\documentclass{ieeeaccess}
\usepackage{cite}
\usepackage{amsmath,amssymb,amsfonts}
\usepackage{algorithmic}
\usepackage{graphicx}
\usepackage{textcomp}
\usepackage{float}
\usepackage{algorithmic}
\usepackage{graphicx}
\usepackage{textcomp}
\usepackage{hhline}
\usepackage{threeparttable}
\usepackage{tabularx}
\usepackage{multirow}
\usepackage{balance}
\usepackage{verbatim}
\usepackage{adjustbox}
\usepackage{array}
\usepackage{makecell}

\newcommand{\etal}{\emph{et~al.}}

\def\BibTeX{{\rm B\kern-.05em{\sc i\kern-.025em b}\kern-.08em
    T\kern-.1667em\lower.7ex\hbox{E}\kern-.125emX}}

\usepackage{tikz}
\NewSpotColorSpace{PANTONE}
\AddSpotColor{PANTONE} {PANTONE3015C} {PANTONE\SpotSpace 3015\SpotSpace C} {1 0.3 0 0.2}
\SetPageColorSpace{PANTONE}%
  
\usepackage{textcomp}
\usepackage{hyperref}
\newcommand\copyrighttext{%
  \footnotesize \textcopyright 2023 IEEE. This article has been accepted for publication in IEEE ACCESS. See \url{http://www.ieee.org/publications_standards/publications/rights/index.html} for copyright information.}
\newcommand\copyrightnotice{%
\begin{tikzpicture}[remember picture,overlay]
\node[anchor=south,yshift=4pt] at (current page.south) {\fbox{\parbox{\dimexpr\textwidth-\fboxsep-\fboxrule\relax}{\copyrighttext}}};
\end{tikzpicture}%
}
    
\begin{document}
\history{Date of publication xxxx 00, 0000, date of current version xxxx 00, 0000.}
\doi{10.1109/ACCESS.2023.3346315} 

\title{ThoraX-PriorNet: A Novel Attention-Based Architecture Using Anatomical Prior Probability Maps for Thoracic Disease Classification}
\author{\uppercase{Md. Iqbal Hossain}\authorrefmark{1, *}, \uppercase{Mohammad Zunaed}\authorrefmark{1, *}, \uppercase{Md. Kawsar Ahmed}\authorrefmark{1}, \uppercase{S.M. Jawwad Hossain}\authorrefmark{1}, \uppercase{Anwarul Hasan}\authorrefmark{2,3} and \uppercase{Taufiq Hasan}\authorrefmark{1,4} \IEEEmembership{Senior Member, IEEE}}
\address[*]{Md. Iqbal Hossain and Mohammad Zunaed contribute equally}
\address[1]{mHealth lab, Department of Biomedical Engineering, Bangladesh University of Engineering \& Technology, Dhaka 1205, Bangladesh}
\address[2]{Department of Mechanical and Industrial Engineering, Qatar University.}
\address[3]{Biomedical Research Center, Qatar University, Doha, Qatar.}
\address[4]{Center for Bioengineering Innovation and Design (CBID), Department of Biomedical Engineering, Johns Hopkins University, Baltimore, MD.}

\markboth{Hossain \headeretal: IEEE ACCESS}{ThoraX-PriorNet: A Novel Attention-Based Architecture Using Anatomical Prior Probability Maps for Thoracic Disease Classification}

\corresp{Corresponding authors: Taufiq Hasan (taufiq@bme.buet.ac.bd) and Anwarul Hasan (ahasan@qu.edu.qa).\\\vspace{3mm}
This work was supported by the Open Access funding provided by the Qatar National Library.}

\begin{abstract}
\textit{Objective:} Computer-aided disease diagnosis and prognosis based on medical images is a rapidly emerging field. Many Convolutional Neural Network (CNN) architectures have been developed by researchers for disease classification and localization from chest X-ray images. It is known that different thoracic disease lesions are more likely to occur in specific anatomical regions compared to others. This article aims to incorporate this disease and region-dependent prior probability distribution within a deep learning framework.
\textit{Methods:} We present the ThoraX-PriorNet, a novel attention-based CNN model for thoracic disease classification. We first estimate a disease-dependent spatial probability, i.e., an \emph{anatomical prior}, that indicates the probability of occurrence of a disease in a specific region in a chest X-ray image. Next, we develop a novel attention-based classification model that combines information from the estimated \emph{anatomical prior} and automatically extracted chest region of interest (ROI) masks to provide attention to the feature maps generated from a deep convolution network. Unlike previous works that utilize various self-attention mechanisms, the proposed method leverages the extracted chest ROI masks along with the probabilistic \emph{anatomical prior} information, which selects the region of interest for different diseases to provide attention. 
\textit{Results:} The proposed method shows superior performance in disease classification on the NIH ChestX-ray14 dataset compared to existing state-of-the-art methods while reaching an area under the ROC curve (\%AUC) of 84.67. Regarding disease localization, the anatomy prior attention method shows competitive performance compared to state-of-the-art methods, achieving an accuracy of 0.80, 0.63, 0.49, 0.33, 0.28, 0.21, and 0.04  with an Intersection over Union (IoU) threshold of 0.1, 0.2, 0.3, 0.4, 0.5, 0.6, and 0.7, respectively. \textit{Impact Statement:} The proposed ThoraX-PriorNet can be generalized to different medical image classification and localization tasks where the probability of occurrence of the lesion is dependent on specific anatomical sites.
\end{abstract}

\begin{keywords}
Anatomical prior, Anatomy-aware attention, Chest X-ray, Thoracic disease classification.
\end{keywords}

\titlepgskip=-15pt

\maketitle

\section{Introduction}
\label{sec:introduction}
\copyrightnotice
Thoracic disorders are one of the major health concerns worldwide as the heart and lungs, two vital human organs,  are located within the thorax. In 2017, around 544·9 million people were affected by chronic respiratory illness \cite{soriano2020prevalence}, a thoracic disease, leading to 3.9 million deaths\cite{bmj}. Various medical imaging modalities, e.g., X-ray, Magnetic Resonance Imaging (MRI), and Computed Tomography (CT) can diagnose different thoracic disorders. The chest X-ray (CXR) remains the most commonly performed and widely available radiological diagnostic method to assess and diagnose thoracic diseases. The chest radiograph is an X-ray projection image of the thoracic cavity used to diagnose conditions affecting the chest, its contents, and nearby structures. It is one of the most effective and low-cost methods for diagnosing thoracic diseases. Since CXR is a projection imaging method providing a 2D image of the 3D thoracic structure, anatomical structures are overlapped in the resulting image. Therefore, diagnosis of diseases with CXR image highly depends on the skill and experience of the radiologist \cite{kelly2016development}. However, in many underserved regions of the world, the number of skilled radiologists is insufficient. In such scenarios, automated CXR image interpretation using artificial intelligence (AI) can significantly benefit health systems. This is true even if the algorithms are not making full autonomous decisions and are only used to assist physicians.\par
\copyrightnotice
However, it is of paramount importance for the machine learning models to be explainable for the radiologists to trust them. Thus, providing an accurate location for the predicted pathologies is a prerequisite for computer-aided diagnosis. However, due to the lack of pixel-level ground truth annotation data, the deep learning models suffer from sub-optimal optimizations. A number of weakly supervised disease localization methods over the recent years have been proposed to solve this problem. In the literature, different attention-based approaches \cite{cai2018iterative,chen2019lesion,ouyang2020learning} have been used for medical disease diagnosis, where the model traditionally learns to identify and focus on the regions of interest containing the lesions using activated feature maps from the classifiers. However, these methods are data-driven and are generally agnostic to the human anatomy and its dependence on identifying the diseased regions. They do not take into account the typical occurrence areas for a specific pathology, and thus, they often fail to predict the lesion region as recognized by radiologists. Intuitively, radiologists do not search all the parts when diagnosing chest X-ray images of a patient for thoracic diseases. Instead, they concentrate on the areas related to the symptoms of the disease of a patient.\par
Different thoracic disease lesions have unique characteristics and are identified in specific regions of a chest radiographic image. For example, when identifying pneumonia, a radiologist looks for white spots in the lungs that show the characteristics of infection. In contrast, the opacity features of pleural effusion manifest in the pleural space, not inside the lung region. Similarly, the cardiomegaly pathology is associated with the heart. Thus, we may consider that the diagnostic features of different thoracic diseases have a higher probability of occurrence in certain anatomical regions of the chest X-ray. Consequently, specific disease features may have a zero probability of occurrence in certain anatomical regions (e.g., observing consolidation features outside the lungs). Therefore, to reliably detect and localize thoracic diseases, we not only require deep learning-based models to learn the disease-specific features but also to focus on the specific anatomical regions where the likelihood of the disease is highest. However, the existing literature studies predict only the most discriminative areas for the pathology localization and classification of a patient without considering the prior distribution knowledge of the regions where a pathology most repeatedly appears. Although Chen \etal \cite{chen2020two} and Kamal \etal \cite{kamal2022anatomy} utilized lung segmentation-based attention mechanisms, disease-specific anatomical prior knowledge was not considered within the attention mechanism and abnormality localization.\par
Considering the limitations of previous works in this area, we propose a novel model architecture using two types of attentions: chest region of interest mask-based attention and disease-specific anomaly-based attention for disease classification. The main contributions of this paper are as follows:
\begin{itemize}
\item We propose the concept of a novel probabilistic anatomical prior map that provides a spatial probability distribution of a disease occurrence within X-ray images. To the best of our knowledge, the idea of a disease-specific anatomical prior probability maps generated using an aggregation of disease ROI masks has not been explored in previous research works.
\item We developed an end-to-end model ThoraX-PriorNet, a novel attention-based architecture that focuses on specific regions of an X-ray image informed by both disease-specific anatomical prior probability maps and lung region-of-interest (ROI) masks.
\item We conducted a thorough experimental evaluation to compare the performance of the proposed ThoraX-PriorNet model with the existing methods. Detailed ablation studies conducted using the anatomy prior attention module (APAM) demonstrate the effectiveness of the proposed method in accurately detecting thoracic diseases.
\end{itemize}
The rest of our document is organized as follows. Section \ref{sec:Related Works} reviews the related works in the thoracic disease classification and weakly supervised localization tasks. Section \ref{sec:Proposed Method} presents our proposed approach in detail. Section \ref{sec:Implementational Details} discusses our experimental settings, such as datasets, data preparation, training scheme, and so on. We conduct comprehensive experiments in Section \ref{sec:Experimental Results}, including ablation studies, performance comparison with state-of-the-art methods, statistical analysis, and so on, both for classification and localization tasks. In section \ref{sec:Conclusion}, we conclude this paper.
\Figure[!t]()[width=\linewidth]{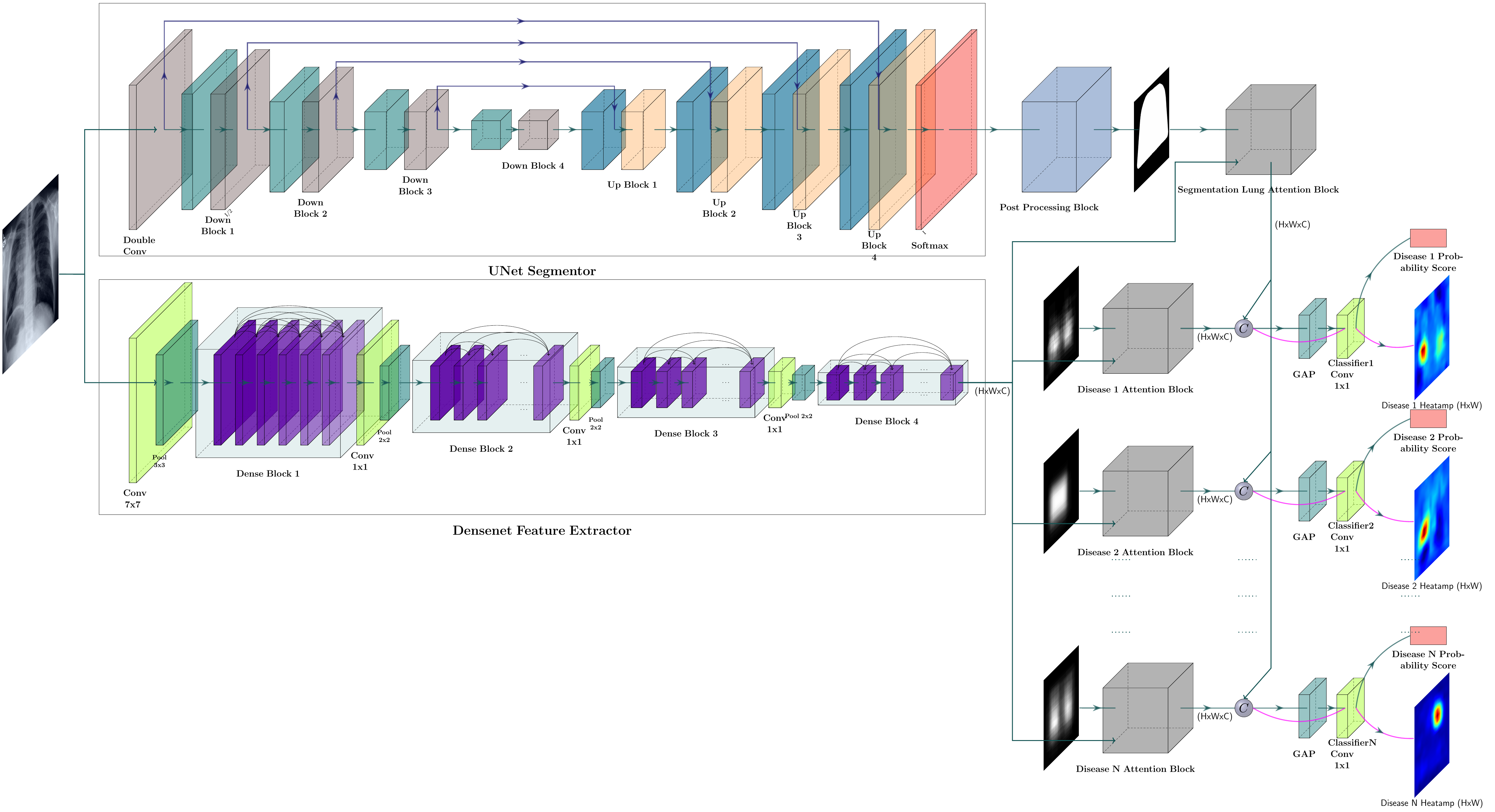}
   {A schematic of the proposed ThoraX-PriorNet architecture for disease classification from CXR utilizing both lung segmentation attention and disease-specific attention. The model consists of three components: the lung segmentation attention module, the disease-specific attention module, and then concatenation for classification. The lung segmentation U-Net model generates a lung ROI mask, which is then used to provide lung mask guided attention. The disease-specific probability mask is used along feature map to provide disease specific attention. Finally, the concatenated feature maps are used to make the final disease classification.\label{fig:model_lungabnormmask}}

\section{Related Works}
\label{sec:Related Works}
\subsection{Attention}
Attention mechanisms that selectively attend to zones of an image with a high probability of exhibiting particular diseases can yield a substantial performance improvement for machine learning models\cite{obeso2022visual}. Chest X-rays are frequently employed for diagnosing respiratory and cardiovascular conditions, precise interpretation of these images is imperative for effective treatment \cite{ullah2023densely, innat2023convolutional, hong2021multi, guendel2018learning , ye2020weakly, baltruschat2019comparison}. Attention modules available in the computer vision literature can be divided into two main categories. One includes the Squeeze-and-Excitation (SE) approach that adaptively re-calibrates channel-wise feature responses by explicitly modeling inter-dependencies between channels \cite{Hu2020Squeeze-and-ExcitationNetworks}. The other is Gather-Excite (GE) method which efficiently aggregates feature responses from a large spatial extent and excites, redistributing the pooled information to local features \cite{Hu2018Gather-Excite:Networks}. Chen \etal \cite{chen2020nonlocal} presented a non-local (NL) attention module to utilize the local relationship for capturing long-range dependencies. Wang \etal \cite{wang2021triple} have introduced a triplet attention model that can learn channel-wise, element-wise, and scale-wise attention simultaneously. This approach helps to capture distinctive information relevant to the task of classifying thorax diseases. Ullah \etal\cite{ullah2023densely} incorporated channel-wise attention as layer in multiple positions in the their feed forward network for Covid-19 classification. Zhang \etal \cite{zhang2021part} presented attention guided with different parts of lung. Kamal \etal \cite{kamal2022anatomy} used lung segmentation mask to provide attention in the lung region in a chest X-ray image. To overcome the domain mismatch of lung segmentation dataset they used GAN model to segment lung that was later used for providing attention.\par
Though providing attention modules in network enhances model performance, most existing approaches mainly focus on learning the attention map using global CXR images, without considering disease specific lung regions. Aiming to address this constraint, the proposed method generated disease specific probabilistic map from the provided bounding box annotation. Then, we provided probabilistic map guided and lung mask guided attention to focus at specific regions in chest X-ray image for thoracic disease evaluation.
\copyrightnotice
\subsection{Weakly Supervised Learning}
Achieving success in supervised learning demands sophisticated network engineering and an enormous quantity of precisely labeled training data \cite{diagnostics12102549}. Weakly supervised learning is becoming increasingly important in medical chest X-ray analysis as it can alleviate the need of extensive and precise annotations required for supervised learning. Wang \etal \cite{wang2017chestx} introduce the ChestX-ray14 dataset, together with a baseline for evaluating weakly supervised lesion localization. Furthermore, numerous studies have previously investigated disease localization on CXR images \cite{wang2017chestx, tang2018attention, yao2018weakly}, without directly utilizing ROI labels. Notably, prior research on localization such as Ye \etal's\cite{ye2020weakly} use of probabilistic-CAM Pooling and Ouyang \etal's \cite{Ouyang2021LearningDiagnosis}  use of hierarchical attention for weakly supervised abnormality localization have incorporated attention mechanisms in their architectures. In their study, Ullah \etal\cite{ullah2023densely} utilized grad-CAM to produce a COVID-19 heatmap, with the aim of presenting classification outcomes that are supported by clinical evidence, and thus applicable to clinical practice. Employing saliency techniques, such as Class Activation Mapping (CAM), Grad-CAM\cite{Selvaraju2016Grad-CAM:Localization}, Grad-CAM++\cite{chattopadhay2018grad}, Eigen-CAM\cite{bany2021eigen}, and similar methods, to produce heatmaps can prove to be highly beneficial in furnishing clinical evidence. E. Rozenberg \etal \cite{Rozenberg2021} achieved high localization performance in regimes by learning to localize the areas with a limited annotation derived from a small fraction masked. Zhu \etal \cite{Zhu2022PCAN} proposed a convolutional attention-based network named PCAN that is pathology-aware and capable of capturing the variations in lesion size and location by generating pixel-wise diagnoses and pixel-wise weights. Han \etal \cite{yan2021Cross} leverage two views, i.e., radiomic and global image features, for training the framework for classifying and localizing thoracic diseases. To extract the radiomic features, they have exploited Grad-CAM generated by the image classifier backbone through a feedback loop mechanism. Xiao \etal \cite{Xiao2022DelvingIM} improved the performance of ViTs by pre-training with 266,340 chest X-rays using Masked Autoencoders, reconstructing missing pixels from a small part of each image. Li \etal \cite{Li2022Model} utilized an adaptive ViT with a DenseNet architecture with a feature pyramid structure to design the inter-patch and patch-wise long-range dependencies and obtain fine-grained feature maps.\par
However, the previous methods from the literature depend on the discriminative power of deep-learning convolutional networks and predict the area of a chest X-ray that is most responsible for classification as lesion area without considering the prior knowledge of the distribution of disease occurrence area in a chest X-ray image.  Instead, we developed an end-to-end novel attention-based architecture named ThoraX-PriorNet, which focuses on specific regions of a chest X-ray image guided by typical disease-specific spatial anatomical prior probability maps.
\copyrightnotice
\section{Proposed Method}
\label{sec:Proposed Method}
This section describes our proposed approach, where we have used a deep learning-based novel classification architecture, named  ThoraX-PriorNet, that utilizes both the chest ROI mask and a disease-specific anatomical prior probability map for pathology classification and localization. We also describe in detail the extraction of the chest ROI mask and the generation of a disease-specific anatomical prior probability map.
\subsection{Generating Disease-specific Anatomical Prior Probability Map}
We compute the disease-specific anatomical prior probability maps by identifying the spatial regions of the CXR images where the lesions are most likely to occur. To construct this map, we use the NIH Chest X-ray dataset, which includes 880 bounding-box annotated images identifying the regions of the abnormality \cite{wang2017chestx}. First, we create a binary image keeping the bounding-box interior spatial values equal to 1 and the rest equal to 0 for a particular disease. Out of the eight pathologies, seven pathologies (atelectasis, effusion, infiltrate, mass, nodule, pneumonia, and pneumothorax) can occur symmetrically in the lungs. Leveraging this behavior, we apply horizontal flipping to bounding boxes of these seven types of diseases to generate new annotations. We then take the sum of all binary images of a particular disease to generate unnormalized probability map. Finally, we normalize pixel values of the unnormalized probability map by dividing them by the maximum pixel value within that probability map. The normalized mask is used in the network as the anatomical prior probability map for providing disease-specific attention.\par
\begin{figure}[!t]
\centerline{\includegraphics[width=\columnwidth]{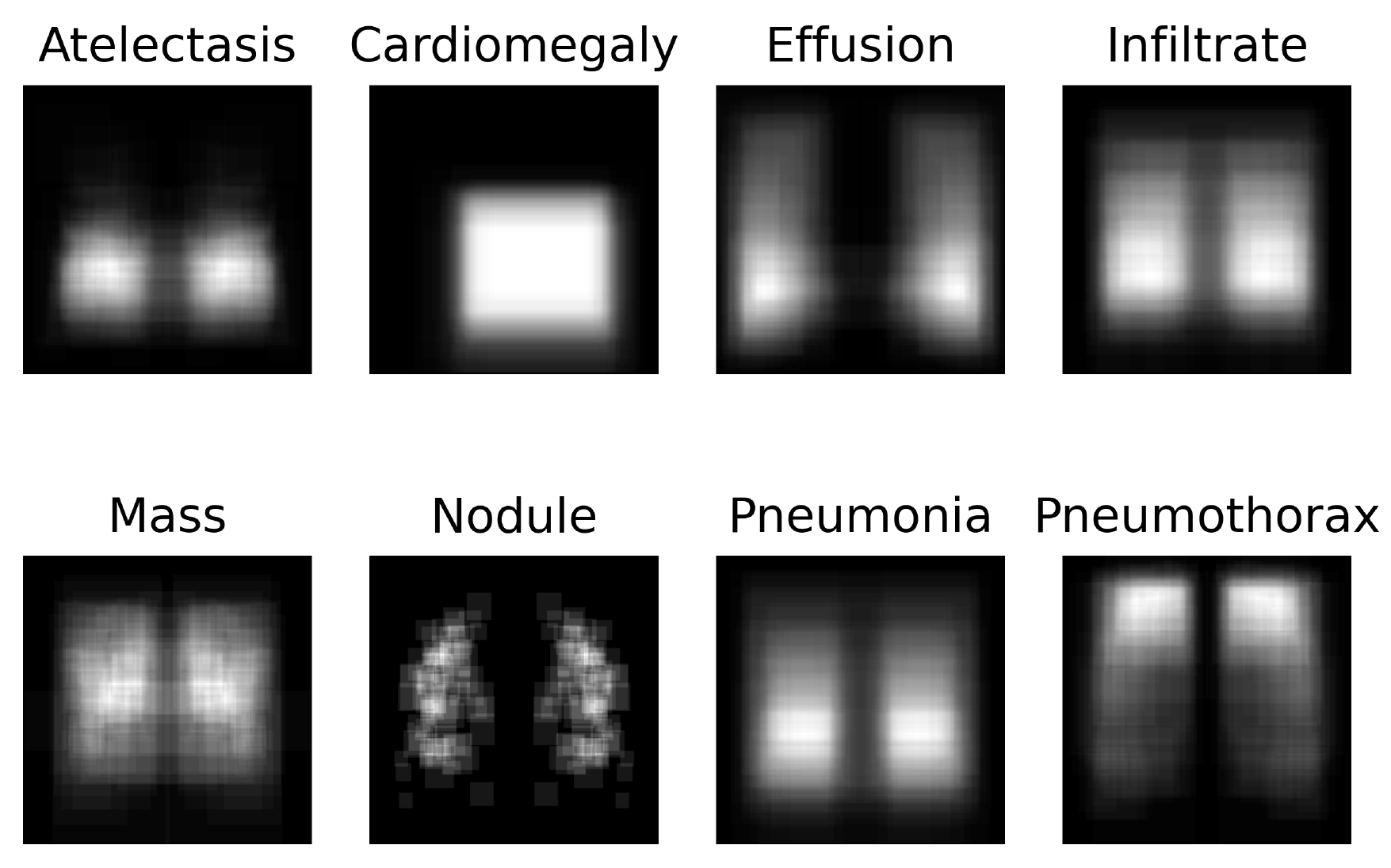}}
\caption{Disease-specific anatomical prior probability maps generated for the 8 diseases for which the bounding box annotations are available in the NIH dataset.}
\label{fig:abnorm_mask}
\end{figure}

First, we obtain the unnormalized raw probability map. Let $I_{c}^{k}(i,j)$ indicate the pixel position ($i$,$j$) of the $k^{th}$ constructed binary mask image from the bounding box annotated ground truth image for the disease class $c$. The disease-specific anatomical prior probability map $\mathit{M}_c^p$ is generated as follows. 
\begin{equation}
\mathit{\hat{M}}_c(i,j) = \sum_{k=1}^{N_c} I_{c}^{k}(i,j) \label{eq7}
\end{equation}
where $N_c$ indicates the number of CXR images available for the disease class $c$. Next, we normalize the raw map $\mathit{\hat{\mathit{M}}}_c$ to obtain the final anatomical prior probability map by,
\begin{equation}
\mathit{M}_c^p(i,j) = \frac{\hat{M}_c(i,j)}{\max \left(\mathit{\hat{M}_c}\right)} \label{eq8} 
\end{equation}
Here, the max operation identifies the maximum pixel value of the raw probability map $\hat{\mathit{M}}_c$. 
Finally, these anatomical prior probability maps were generated for all eight diseases for which the bounding box annotations are available. Fig.~\ref{fig:abnorm_mask} shows the generated disease-specific anatomical prior probability maps for the eight abnormalities. In the strictest sense, the obtained maps $\mathit{M}_c^p(i,j)$ do not represent an actual probability distribution. Firstly, this is because the regions are obtained from the bounding box information that is larger than the actual disease regions. Secondly, obtaining a probability distribution requires that the integration over the entire map should equal unity. In actual implementation, the map's relative intensity values are more important than the absolute values. For, disease classes whose bounding-box annotations are not available, we used ${M}_c^p(i,j)=1$.

\subsection{Chest ROI Mask Generation}
\begin{figure}[!t]
\centering
\includegraphics[width=\columnwidth]{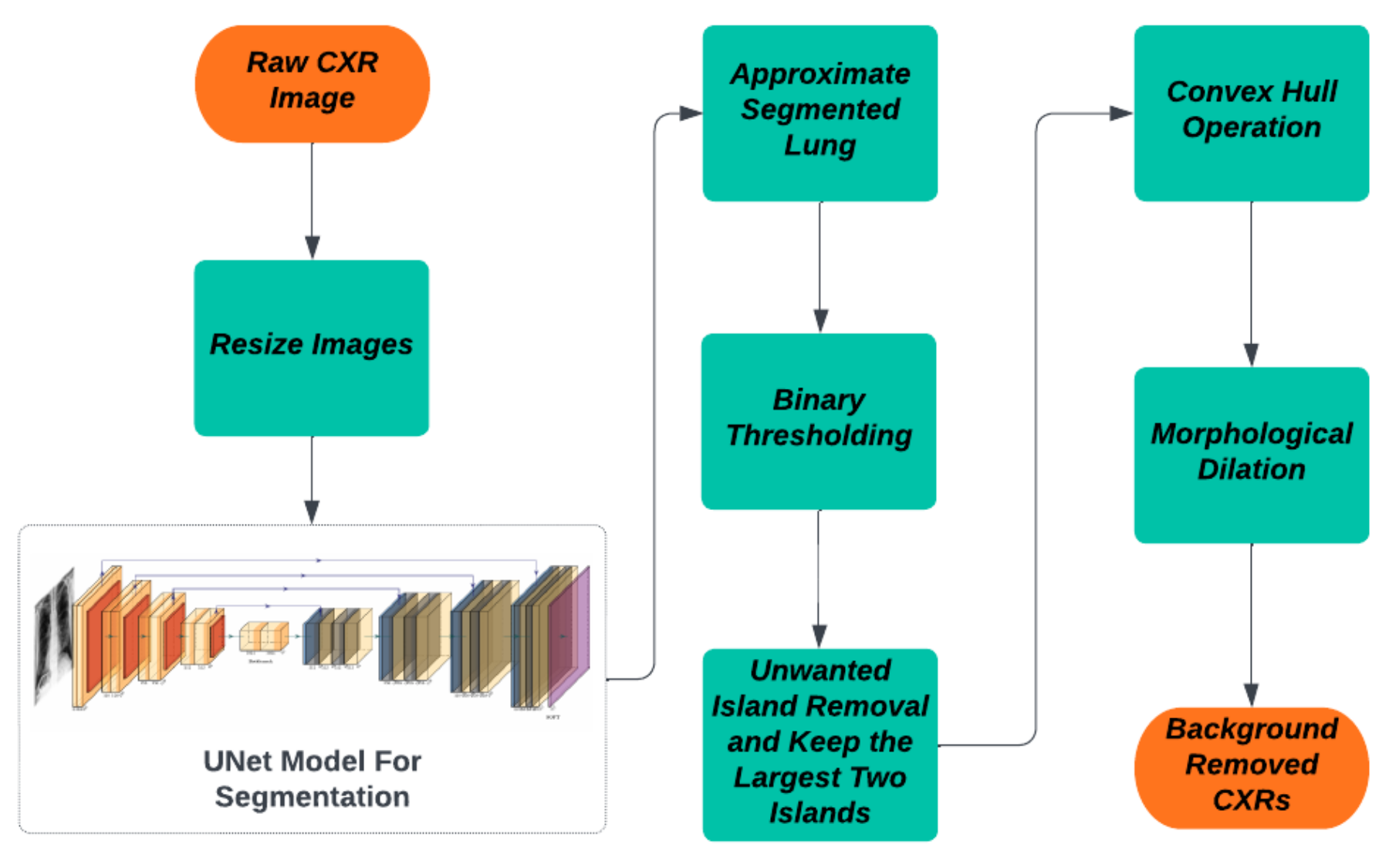}
\caption{A flow-diagram of the chest ROI mask generation module.}
\label{fig:segworkflow}
\end{figure}

\begin{figure}[!t]
\centering
\includegraphics[width=\columnwidth]{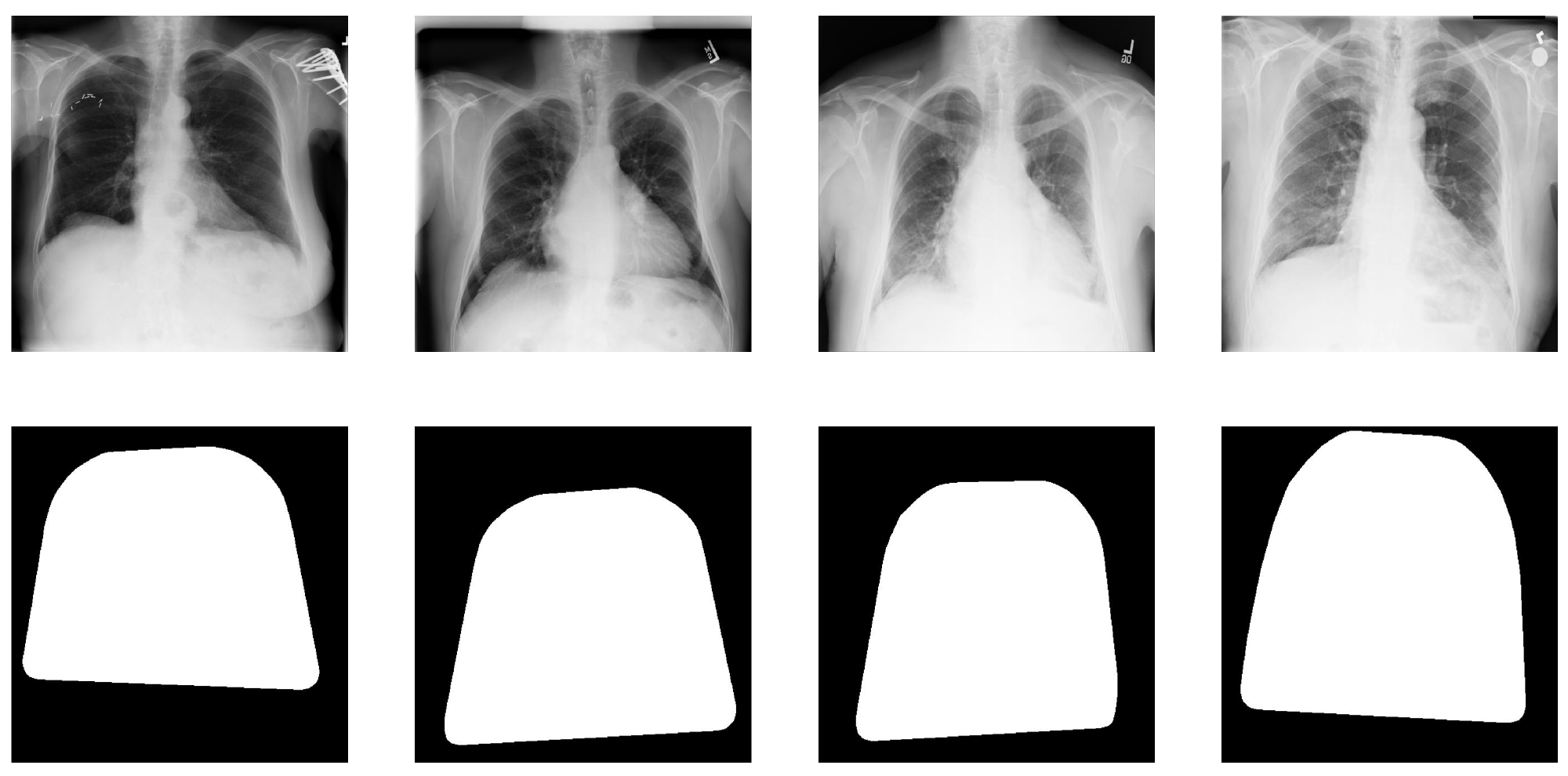}
\caption{Examples of some generated chest ROI masks. Top panel:
Example input CXR images, Bottom panel: Corresponding chest ROI masks of
the example CXR images.}\label{fig:segfig}
\end{figure}

We employ the well-established U-net \cite{Ronneberger2015U-Net:Segmentation} segmentation model to extract the lung regions from the input CXR images. We train the model using the 247 images from the JSRT dataset \cite{shiraishi2000development}. The segmentation model produces undesirable small islands in the case of some images. To address these issues, we binarize and apply post-processing to the segmentation results to remove the unwanted islands based on the anatomical characteristics of the lungs. Since all other islands are small compared to the lung islands, we filter out the largest two islands representing the right and left lung. The sternum region is also important for some thoracic diseases and contains crucial information for classification. To retain this region, we use the convex hull operation \cite{preparata1977convex}. Finally, we use morphological expansion to retain further information from the pleural regions. The overall chest ROI mask generation flow chart is provided in Fig.~\ref{fig:segworkflow}. Some of the CXR images and their corresponding generated masks are shown in Fig.~\ref{fig:segfig}. These postprocessing operations are represented by the postprocessing block in the ThoraX-PriorNet full architecture in Fig.~\ref{fig:model_lungabnormmask}.
\copyrightnotice
\subsection{Anatomical Prior Attention Module (APAM)}
\begin{figure*}[!t]
\centering
\includegraphics[width=\linewidth]{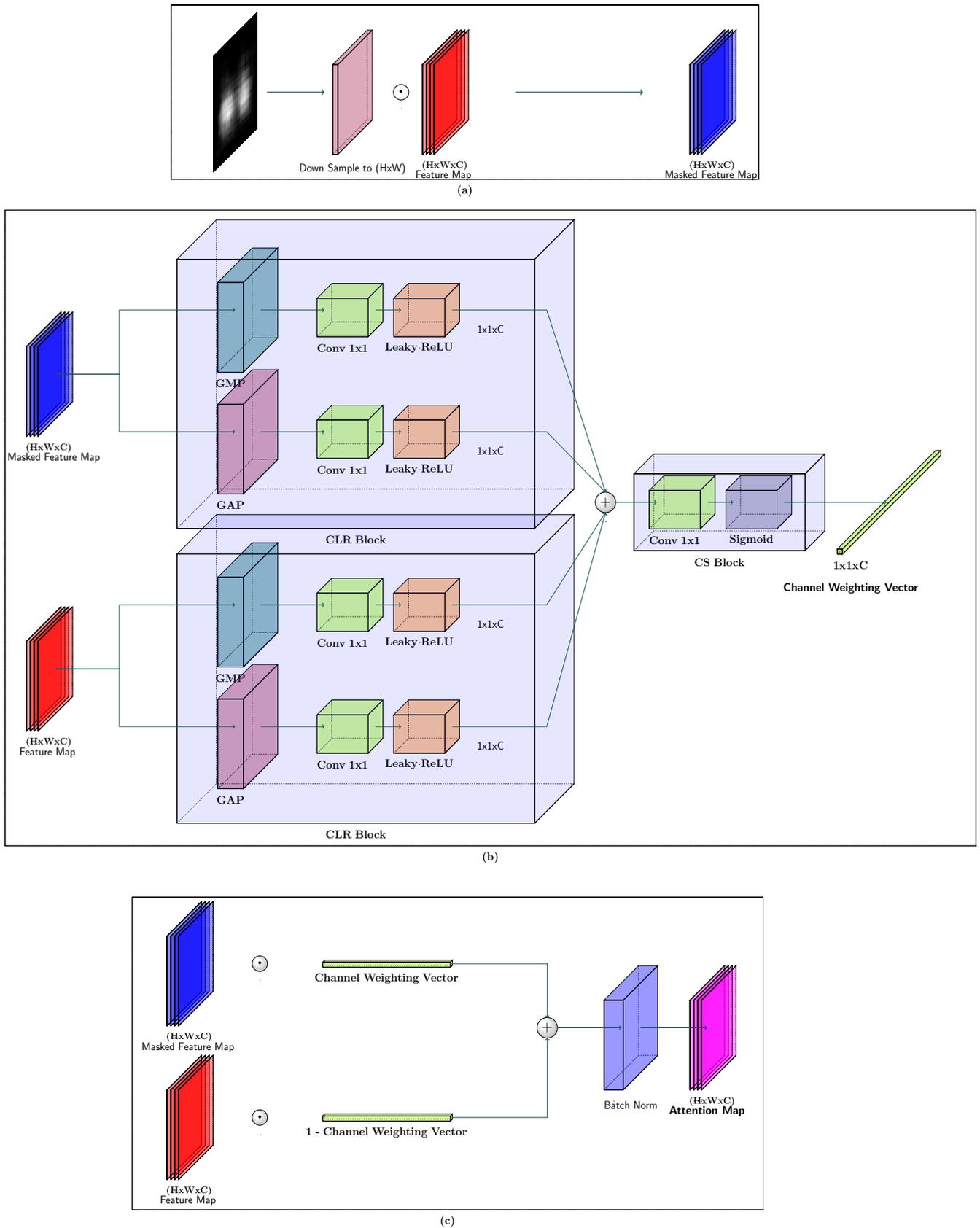}
\caption{\textbf{A schematic diagram of the Anatomy Prior Attention Module (APAM):} A) Mask is multiplied with the feature map to generate a masked feature map; B) Featuremap and Masked Featuremap is being used to produce channel weighting vector. Here, GMP = Global Max Pooling and GAP = Global Average Pooling; C) Channel weighting vector is being used to produce final weighted featuremap i.e, Attention Map.} \label{fig:attention_module}
\end{figure*}

In this section, we describe the anatomical prior attention module (APAM), which takes a feature map and a mask (chest ROI mask or anomaly probability map) as inputs to generate an attention map by providing spatial attention to the feature map. An illustration of the APAM framework is demonstrated in Fig.~\ref{fig:attention_module}. First, we multiply the feature map with the input mask to generate a masked feature map. Later, we take the weighted sum of the feature map and masked feature map to retain information from the region outside the mask since some disease predictions may depend on the feature of the unmasked region. The weights are generated from the feature map and the masked feature map through a CNN. To learn the weights, we use a network similar to the channel-wise attention module described in \cite{woo2018cbam}. However, unlike \cite{woo2018cbam}, we aggregate spatial information from both the feature map and the masked feature map.\par
Let $\mathbf{\mathit{F}}\in\mathbb{R}^{C \times H \times W}$ be the feature map generated by the backbone CNN network and $\mathit{M}_{inp}\in\mathbb{R}^{1 \times H \times W}$ be the input mask (chest ROI mask or anomaly probability map) resized to the spatial dimension of feature map $\mathit{F}$. We pass the feature map $\mathit{F}$ into two pooling layers: global average pooling (AvgPool) and global max pooling (MaxPool). The two corresponding outputs from these pooling layers are denoted as $\mathit{F}_{avg}$ and $\mathit{F}_{max}$ respectively, where $\mathit{F}_{avg}, \mathit{F}_{max}\in\mathbb{R}^{C \times 1 \times 1}$. Again, let $\mathit{F}_{m}\in\mathbb{R}^{C \times H \times W}$ be the masked feature map which is produced after we multiply the feature map $\mathit{F}$ with the input mask $\mathit{M}_{inp}$. We obtain $\mathit{M}_{avg}, \mathit{M}_{max}\in\mathit{R}^{C \times 1 \times 1}$ after passing $\mathit{M}$ through the global average pooling and global max pooling layers in a similar way.
\begin{align}
\mathit{F}_{m} &= \mathit{F} \odot \mathit{M}_{inp} \\
\mathit{F}_{avg} &= \textrm{AvgPool}\left(\mathit{F}\right) \\
\mathit{F}_{max} &= \textrm{MaxPool}\left(\mathit{F}\right) \\
\mathit{M}_{avg} &= \textrm{AvgPool}\left(\mathit{M}\right) \\
\mathit{M}_{max} &= \textrm{MaxPool}\left(\mathit{M}\right) 
\end{align}
Here, $\odot$ denotes element wise multiplication. Furthermore, instead of shared multi-layered perceptron (MLP), we use separate MLPs for all four spatial context descriptors $\left(\mathit{F}_{avg}, \mathit{F}_{max}, \mathit{M}_{avg}, \mathit{M}_{max}\right)$. After passing the spatial context descriptors through the CNN, the network produces the required channel weighting values, $\mathit{W}\in\mathbb{R}^{C \times 1 \times 1}$. The mathematical equation for generating the channel weighting values, $W$ is provided below:
\begin{align}
\mathit{W} = &\textrm{CS} \Bigl( \textrm{CLR\textsubscript{1}}\left(\mathit{F}_{avg}\right) + \textrm{CLR\textsubscript{2}}\left(\mathit{F}_{max}\right) + \nonumber \\  &\textrm{CLR\textsubscript{3}}\left(\mathit{M}_{avg}\right) + \textrm{CLR\textsubscript{4}}\left(\mathit{M}_{max}\right)\Bigr)\label{eq5}
\end{align}
Here,\textrm{CLR\textsubscript{1}},\textrm{CLR\textsubscript{2}}, $\ldots$ ,\textrm{CLR\textsubscript{4}} indicate the blocks of sequential convolutional layer, and leaky ReLU activation layer and then \textrm{CS} indicates block of sequential convolutional layer followed by sigmoid activation layer. In \textrm{CS} block, we use the sigmoid activation function so that the components of weight $\mathit{W}$ are within the range $[0, 1]$. For the \textrm{CLR} blocks, we use the leaky ReLU with a negative slope of 0.2 to mitigate the vanishing gradient problem \cite{Maas2013RectifierModels}. Finally, we generate the attention map $\mathit{A}\in\mathbb{R}^{C \times H \times W}$ from the weighted sum of $\mathit{F}$ and $\mathit{F}_{m}$ using the formula below:
\begin{equation}
\mathit{A} = \mathit{W} \odot \mathit{F} + \left(1 - \mathit{W}\right) \odot \mathit{F}_{m}\label{eq6}
\end{equation}

\subsection{Classification and Localization}
At first, we extract a feature map from the input image with a CNN backbone. We have used DenseNet-121 \cite{Huang2016DenselyNetworks} as backbone for feature extraction. Then we use APAM to generate an attention map from the extracted feature map. For generating attention maps from the feature map, we have used the image-specific chest ROI mask described previously with APAM to generate ROI attention map $\mathit{A}_{ROI}$.
\begin{equation}
\mathit{A}_{ROI} = \mathit{W}_{ROI} \odot \mathit{F} + \left(1 - \mathit{W}_{ROI}\right) \odot \mathit{F}_{ROI}
\end{equation}
Here, $\mathit{W}_{ROI}$ is the weight generated by APAM from feature map $\mathit{F}$ and masked feature map $\mathit{W}_{ROI}$. Then, we have used $K$ ($K$ = number of abnormalities) numbers of disease-specific anatomy prior probability maps with APAM to generate $K$ disease-specific attention maps $\mathit{A}_{p}^c$.
\begin{equation}
\mathit{A}_{p}^c = \mathit{W}_{p}^c \odot \mathit{F} + \left(1 - \mathit{W}_{p}^c\right) \odot \mathit{F}_{p}^c,\ c \in \{1,\dots,K\}
\end{equation}
Here, $\mathit{W}_{p}^c$ is the weight generated by APAM from feature map $\mathit{F}$ and masked feature map $\mathit{F}_{p}^c$ of abnormality $c$. Then for predicting the probability of each disease, the image-specific ROI attention map and the disease-specific attention map of that particular disease are channel-wise concatenated to produce a disease-specific concatenated map. 
\begin{equation}
\mathit{A}_{cat}^c = \textrm{concat}(\mathit{A}_{ROI}, \mathit{A}_{p}^{c}),\ c \in \{1,\dots,K\}
\end{equation}
Here, $\mathit{A}_{cat}^c \in \mathbb{R}^{2C\times H \times W}$. These concatenated maps are passed through individual global pooling and then $1\times1$ convolutional layers sequentially to generate the probability of that disease. And we have used the same convolutional layers on the concatenated feature maps to generate individual heatmap using CAM method. The schematic of proposed architecture of ThoraX-PriorNet is shown in Fig.~\ref{fig:model_lungabnormmask}.

\subsection{Loss Function}
We concatenate the predicted raw values from each of the pathology-specific classifiers and pass them through a sigmoid layer to generate the probabilities, $p^s=[p^s_1, \ldots, p^s_i, \ldots, p^s_c]$. Here, $c$ represents the number of pathologies presented in a dataset. The ground truth vectors of each chest X-ray are expressed as an $c$-dimensional label vector, $L=[l_1, \ldots, l_i, \ldots, l_c]$, where $l_i \in \{0, 1\}$. $l_i$ denotes whether there is any pathology, i.e., 1 for presence and 0 for absence. We optimize the weight parameters of our model by minimizing the binary cross-entropy loss, defined as,
\begin{equation}
\mathcal{L}= -\frac{1}{c}\sum_{i=1}^{c}\bigg[l_i\log{\left(p^s_{i}\right)}+(1-l_i)\log{\left(1-p^s_{i}\right)}\bigg] 
\end{equation}

\section{Implementational Details}
\label{sec:Implementational Details}
\subsection{Data Resources}
We evaluate the proposed ThoraX-PriorNet architecture on the NIH ChestX-Ray14 and CheXpert datasets. These data resources are briefly described below.\par
\textbf{NIH ChestX-Ray14}: The NIH ChestX-Ray14 contains $112,120$ frontal chest X-ray images from 30,805 unique patients \cite{wang2017chestx}. All these images are annotated for 15 classes (14 diseases along with “No Findings”). Within this dataset, 880 images are specially annotated by a bounding box for the localization of 8 diseases. In our classification experiments, we use 70\%, 10\%, and 20\% data for training, cross-validation, and testing, respectively. We train and test our model on the classification data for all 15 classes. On the other hand, we use the bounding-box annotated data of the 8 classes to assess the disease localization performance of our model. Note that there is no patient overlap between all the training, validation, and testing sets. The 880 images with bounding box information are not utilized in training or validation splits.\par
\textbf{CheXpert}: The CheXpert dataset \cite{irvin2019chexpert} is a chest X-ray dataset containing class label annotation of 14 classes (13 diseases along with “No Findings”). Other than positive and negative labels for each class, the dataset also contains an uncertainty label for some images. The dataset consists of 224,316 chest X-ray images for training and 230 chest X-ray images for validation. We use only frontal view chest X-ray images from this dataset. If we consider only images with a frontal view, there are about 200,000 chest X-ray images for training and 200 images for validation in the dataset. We use this dataset for the classification of five thoracic diseases, namely, atelectasis, cardiomegaly, consolidation, edema, and pleural effusion. 
\begin{figure}[!t]
    \centering
    \includegraphics[width=\linewidth]{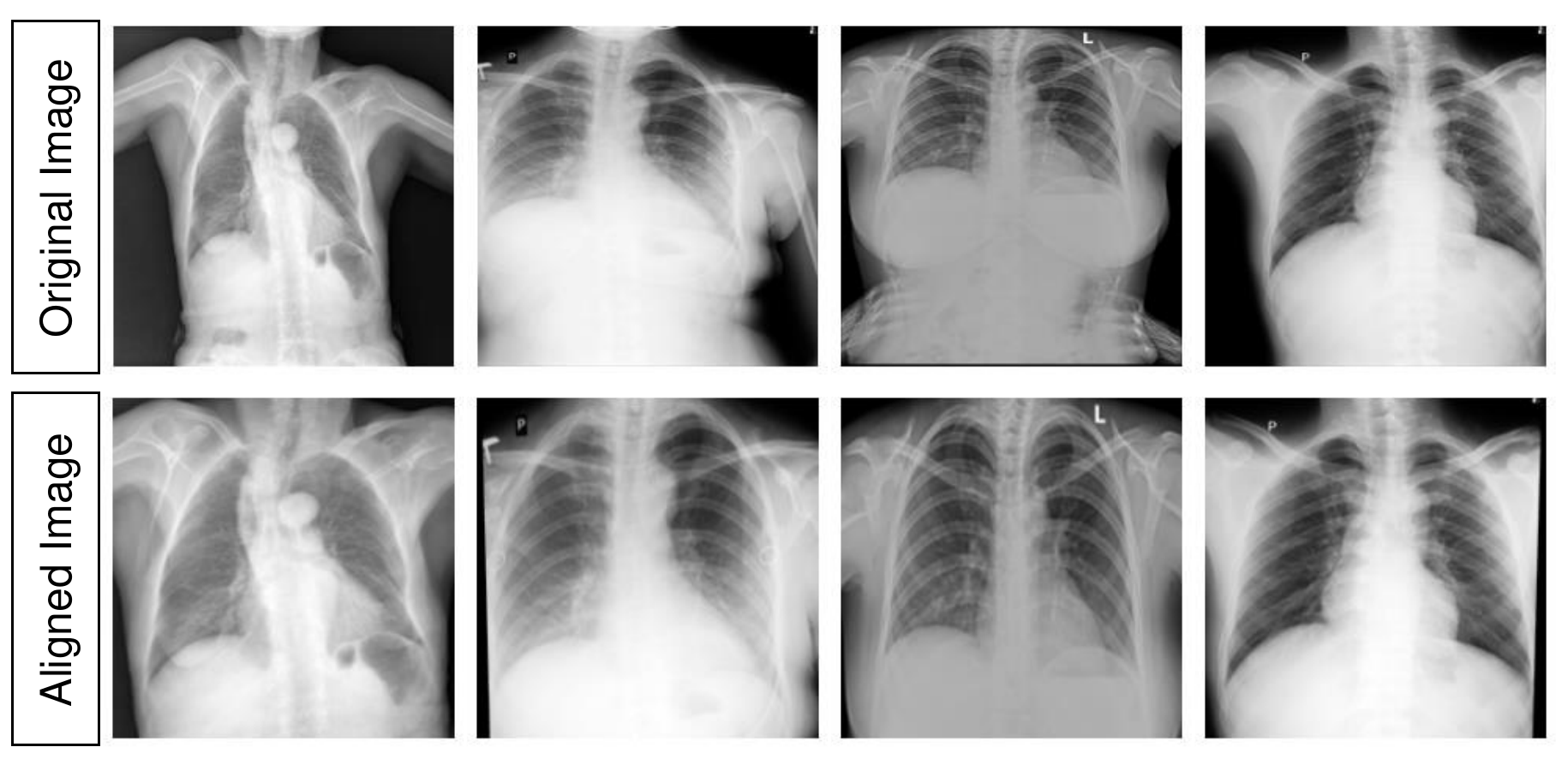}
    \caption{Examples of original chest X-ray images and aligned chest X-ray images.}
    \label{fig: alignment_examples}
\end{figure}
\copyrightnotice
\subsection{Data Preparation}
The chest X-ray images from a dataset generally have diverse variations, such as rotations, shifts, and different scales, making it challenging for the deep-learning models to localize the lesion areas. To address this problem, we utilize the alignment module \cite{liu2019align} to perform spatial alignment on all the images as well as on the bounding box images for generating abnormality masks. Given the input image $I$, the alignment module $\phi$ transforms $I$ to $\phi(I)$. The canonical chest X-ray image, known as the target image $T$, is generated by randomly selecting two thousand normal chest X-ray images and averaging them to a single image. To provide $\phi(I)$ with an aligned structure, we minimize the feature reconstruction loss \cite{10.1007/978-3-319-46475-6_43} between $\phi(I)$ and $T$. The backbone of the alignment module consists of ResNet-18 architecture. The output of the alignment network is the affine transformation parameters. Finally, the affine transformation is applied to the original chest X-ray images to generate aligned chest X-ray images. Fig. \ref{fig: alignment_examples} shows some examples of original and aligned X-ray images.\par
We first normalize the pixel values of chest X-ray images with the mean and standard deviation of pixels from the ImageNet dataset \cite{Deng2009ImageNet:Database}. Next, we resize the image to $586 \times 586$ pixels. Afterward, the training images are randomly cropped to $512\times512$ pixels \cite{yan2018weakly, Zhu2022PCAN}. The validation and test images are center-cropped to $512\times512$ pixels. We use the same resizing and cropping method for the corresponding anatomy prior maps and chest ROI masks. Following \cite{luo2020deep, yan2018weakly}, we use test time augmentation by utilizing average probabilities of ten cropped sub-images (four corner crops and one central crop and the horizontally flipped version of them) as the final prediction. In the case of CheXpert dataset preparation (image augmentation, dealing with class imbalance, uncertain labels, etc.), we use the same procedure described in \cite{ye2020weakly}. We use the same disease-specific anatomy prior maps computed from the NIH dataset for the CheXpert dataset.

\subsection{Training Parameters}
The Table~\ref{tab:hyperparameters} shows the hyperparameters used for training and evaluation of the deep learning model. These include the number of epochs, batch size, loss function, optimizer, learning rate, learning rate scheduler, and weight decay rate. We have utilized the exponential moving average scheme with an alpha rate of 0.997 for updating the model weight. In addition, we have performed gradient accumulation with a step of eight iterations.

\subsection{Activation Map and Bounding Box Generation}
We use class activation maps (CAM) for heatmap generation. For the generation of bounding boxes from the heatmap map, to evaluate localization performance, we first convert the activation map or heatmap to a binary mask using binary thresholding with a threshold value of 127. Next, we use the algorithm introduced by \cite{Suzuki1985TopologicalFollowing} to find the contours of the regions inside the binary mask and prepare bounding boxes around the contours by taking extreme boundary values of the contours as the edge of our bounding boxes.

\subsection{Evaluation Metrics}
We use ROC-AUC (Receiver Operating Characteristic-Area Under Curve), also abbreviated as AUC, to measure the classification performance of our model on the NIH test data. Furthermore, we use the ratio of the number of cases with correct localization against the total number of cases in each class to report the localization performance of our models on 880 bounding-box annotated data of the NIH dataset. Here, we use IoU (Intersection over Union) between the predicted bounding box and ground-truth to detect correct localization following prior work \cite{Ouyang2021LearningDiagnosis, Li2017ThoracicSupervision, wang2017chestx}. In this case, the localization result is regarded as correct if $IoU > T(IoU)$ where $T(IoU)$ is the threshold for localization.\par
We have chosen the model with the highest AUC score on the validation split for inference on the test dataset. The loss and AUC curves during training and validation on the NIH ChestX-Ray14 dataset are given in Fig.~\ref{fig: auc_loss_curves}.
\begin{table}[!t]
    \caption{Hyperparameters of the deep learning model used for training and evaluation.}
    \centering
    \begin{tabularx}{\columnwidth}{|X|X|}
    \hline
         \textbf{Name of Variable} & \textbf{Values}\\
    \hline
    \hhline{==}
         Epochs & 50 \\
    \hline
        Batch size & 16 \\
    \hline
        Loss Function & Binary cross entropy loss \\
    \hline
        Optimizer & Adam \\
    \hline
        Learning rate & 0.0001 \\
    \hline
        Learning rate scheduler & Exponential, 0.75 per 4 epochs \\
    \hline
    \end{tabularx}
    \label{tab:hyperparameters}
\end{table}
\Figure[!t]()[width=0.4\textwidth]{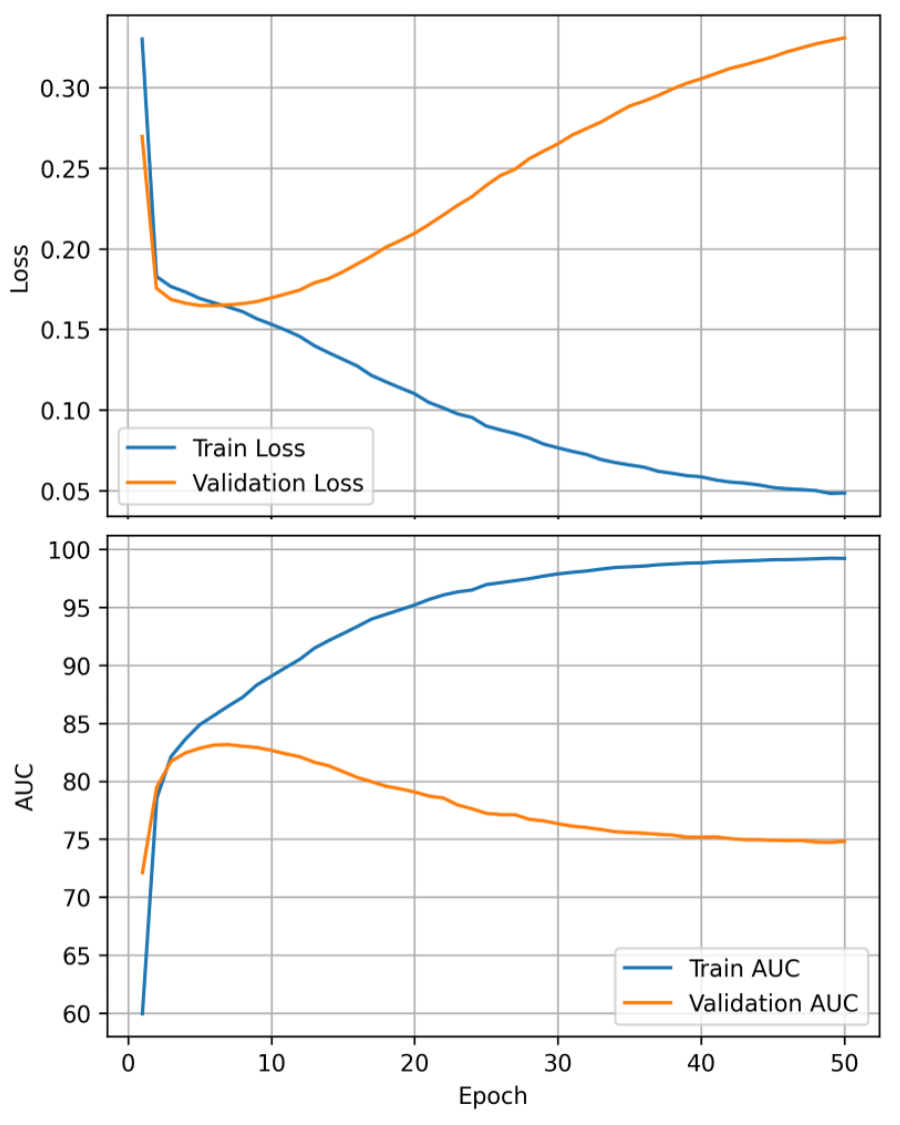}
   {Illustration of the training and validation loss and AUC curves on the NIH ChestX-Ray14 dataset.\label{fig: auc_loss_curves}}

\begin{table*}[!t]
\caption{Ablation Study: Impact of different types of attention masks on the AUC (\%) scores of our trained models on the NIH dataset. The best results are shown in \textcolor{red}{red} font.}
\label{table: classification abl comp}
\centering
\begin{adjustbox}{width=\textwidth}
\begin{threeparttable}
\begin{tabular}{| c | c | c | c | c | c | c | c | c | c | c | c | c | c | c | c | c | c |}
\hline
Method & AM (AbM) & AM (LM) & Atel & Card & Effu & Infil & Mass & Nodu & Pne1 & Pne2 & Cons & Edem & Emph & Fib & PT & Her & Mean \\
\hline
Baseline &   &   & \textcolor{red}{\bf 82.98} & 90.20 & 88.25 & 72.32 & 86.50 & \textcolor{red}{\bf 80.86} & 75.99 & 88.99 & \textcolor{red}{\bf 81.50} & \textcolor{red}{\bf 90.80} & 92.56 & 81.88 & 80.58 & 86.91 & 84.30 \\
\hline
ThoraX- PriorNet &   & \checkmark & 82.61 & 89.78 & 88.25 & 72.30 & 86.83 & 80.76 & 75.48 & \textcolor{red}{\bf 89.35} & 81.04 & 90.50 & \textcolor{red}{\bf 92.85} & 81.99 & 81.00 & 88.16 & 84.35 \\
\hline
ThoraX- PriorNet & \checkmark &   & 82.47 & \textcolor{red}{\bf 90.58} & 88.20 & 72.24 & \textcolor{red}{\bf 86.89} & 80.64 & \textcolor{red}{\bf 76.41} & 88.74 & 81.30 & 90.67 & 92.82 & 82.33 & 80.64 & \textcolor{red}{\bf 91.72} & \textcolor{red}{\textbf{84.69}} \\
\hline
ThoraX- PriorNet & \checkmark & \checkmark & 82.68 & 90.16 & \textcolor{red}{\bf 88.35} & \textcolor{red}{\bf 72.34} & 86.73 & 80.70 & 76.38 & 88.98 & 81.16 & 90.78 & 92.70 & \textcolor{red}{\bf 82.56} & \textcolor{red}{\bf 81.29} & 90.53 & 84.67 \\
\hline
\end{tabular}
\begin{tablenotes}
    \item Here, AM (AbM) = APAM Utilizing Probabilistic Abnormality Mask,  AM (LM) = APAM Utilizing Chest ROI Mask, Atel = Atelectasis, Card = Cardiomegaly, Effu = Effusion, Infi = Infiltration, Nodu = Nodule,  Pne1 = Pneumonia, Pne2 = Pneumothorax, Cons = Consolidation, Edem = Edema, Emph = Emphysema, Fibr = Fibrosis, PT = Pleural Thickening, Her = Hernia
\end{tablenotes}
\end{threeparttable}    
\end{adjustbox}
\end{table*}

\begin{table*}[!t]
\caption{Ablation Study: Impact of input image spatial resolution on the AUC (\%) scores of our trained models on the NIH dataset. The best results are shown in \textcolor{red}{red} font.}
\label{table: classification abl spatial resolution}
\centering
\begin{adjustbox}{width=\textwidth}
\begin{threeparttable}
\begin{tabular}{| c | c | c | c | c | c | c | c | c | c | c | c | c | c | c | c |}
\hline
Method & Atel & Card & Effu & Infil & Mass & Nodu & Pne1 & Pne2 & Cons & Edem & Emph & Fib & PT & Her & Mean \\
\hline
224x224 & 82.54 & 90.57 & \textcolor{red}{\bf 88.35} & 72.29 & 86.39 & 78.01 & \textcolor{red}{\bf 77.00} & 87.96 & \textcolor{red}{\bf 81.89} & \textcolor{red}{\bf 90.98} & 92.38 & 81.75 & 80.04 & \textcolor{red}{\bf 91.90} & 84.43 \\
\hline
368x368 & \textcolor{red}{\bf 82.92} & \textcolor{red}{\bf 90.75} & 88.30 & 72.23 & \textcolor{red}{\bf 86.96} & 80.03 & 76.82 & 88.18 & 81.30 & 90.66 & \textcolor{red}{\bf 92.96} & 82.37 & 80.59 & 90.93 & 84.64 \\
\hline
512x512 & 82.68 & 90.16 & \textcolor{red}{\bf 88.35} & \textcolor{red}{\bf 72.34} & 86.73 & \textcolor{red}{\bf 80.70} & 76.38 & \textcolor{red}{\bf 88.98} & 81.16 & 90.78 & 92.70 & \textcolor{red}{\bf 82.56} & \textcolor{red}{\bf 81.29} & 90.53 & \textcolor{red}{\textbf{84.67}} \\
\hline
\end{tabular}
\begin{tablenotes}
    \item Here, Atel = Atelectasis, Card = Cardiomegaly, Effu = Effusion, Infi = Infiltration, Nodu = Nodule,  Pne1 = Pneumonia, Pne2 = Pneumothorax, Cons = Consolidation, Edem = Edema, Emph = Emphysema, Fibr = Fibrosis, PT = Pleural Thickening, Her = Hernia
\end{tablenotes}
\end{threeparttable}    
\end{adjustbox}
\end{table*}

\begin{table*}[!t]
\caption{Ablation Study: Effect of resizing feature and anatomy prior maps on the AUC (\%) scores of our trained models on the NIH dataset. The best results are shown in \textcolor{red}{red} font.}
\label{table: classification abl feature map resizing}
\centering
\begin{adjustbox}{width=\textwidth}
\begin{threeparttable}
\begin{tabular}{| c | c | c | c | c | c | c | c | c | c | c | c | c | c | c | c |}
\hline
Method & Atel & Card & Effu & Infil & Mass & Nodu & Pne1 & Pne2 & Cons & Edem & Emph & Fib & PT & Her & Mean \\
\hline
16x16 & 82.56 & 90.28 & \textcolor{red}{\textbf{88.15}} & 71.81 & \textcolor{red}{\textbf{87.12}} & 80.50 & 76.24 & \textcolor{red}{\textbf{89.06}} & 80.89 & 90.49 & \textcolor{red}{\textbf{93.07}} & \textcolor{red}{\textbf{83.35}} & 81.02 & \textcolor{red}{\textbf{90.85}} & \textcolor{red}{\textbf{84.66}} \\
\hline
32x32 & 82.42 & 90.47 & 88.01 & 71.78 & 86.71 & 80.38 & 76.40 & 88.89 & 80.77 & 90.69 & 92.75 & 83.24 & 80.69 & 89.21 & 84.46 \\
\hline
48x48 & \textcolor{red}{\textbf{82.96}} & \textcolor{red}{\textbf{90.52}} & 88.14 & \textcolor{red}{\textbf{72.06}} & 86.76 & \textcolor{red}{\textbf{80.82}} & \textcolor{red}{\textbf{76.46}} & 89.00 & \textcolor{red}{\textbf{81.01}} & \textcolor{red}{\textbf{91.09}} & 93.03 & 81.78 & \textcolor{red}{\textbf{81.05}} & 90.29 & 84.64 \\
\hline
\end{tabular}
\begin{tablenotes}
    \item Here, Atel = Atelectasis, Card = Cardiomegaly, Effu = Effusion, Infi = Infiltration, Nodu = Nodule,  Pne1 = Pneumonia, Pne2 = Pneumothorax, Cons = Consolidation, Edem = Edema, Emph = Emphysema, Fibr = Fibrosis, PT = Pleural Thickening, Her = Hernia
\end{tablenotes}
\end{threeparttable}    
\end{adjustbox}
\end{table*}

\begin{table*}[!t]
\caption{Comparison of AUC (\%) Scores of our best performing model with state-of-the-art methods on the NIH dataset. The best results are shown in \textcolor{red}{red} font.}
\label{table:auc_compare}
\centering
\tabcolsep=5.1pt\relax
    \begin{threeparttable}
\begin{tabularx}{\textwidth}{|c|c c c c c c c c c c c c c c|c|}
\hline
Model & Atel & Card & Effu & Infi & Mass & Nodu & Pne1 & Pne2 & Cons & Edem & Emph & Fibr & PT & Hern & Mean \\ \hhline{================}
LSTM-Net\cite{yao2017learning} & 73.30 & 85.80 & 80.60 & 67.50 & 72.70 & 77.80 & 69.00 & 80.50 & 71.70 & 80.60 & 84.20 & 75.70 & 72.40 & 82.40 & 76.73 \\ \hline
TieNet\cite{wang2018tienet} & 73.20 & 84.40 & 79.30 & 66.60 & 72.50 & 68.50 & 72.00 & 84.70 & 70.10 & 82.90 & 86.50 & 79.60 & 73.50 & 87.60 & 77.24 \\ \hline
AGCL\cite{tang2018attention} & 75.57 & 88.65 & 81.91 & 68.92 & 81.36 & 75.45 & 72.92 & 84.99 & 72.83 & 84.75 & 90.75 & 81.79 & 76.47 & 87.47 & 80.27 \\ \hline
Ho \etal\cite{khanh2019multiple} & 79.50 & 88.70 & 87.50 & 70.30 & 83.50 & 71.60 & 74.20 & 86.30 & 78.60 & 89.20 & 87.50 & 75.60 & 77.40 & 83.60 & 80.96 \\ \hline
CARL\cite{guan2020multi} & 78.10 & 88.00 & 82.90 & 70.20 & 83.40 & 77.30 & 72.90 & 85.70 & 75.40 & 85.00 & 90.80 & 83.00 & 77.80 & 91.70 & 81.59 \\ \hline
Liu \etal\cite{liu2022acpl} &79.80 & 89.03 & 83.56 & 71.40 &82.49 & 77.73 & 73.86 & 86.95 & 75.50 & 84.95 & 93.36 & 81.86 & 77.60 & 85.89 & 81.77 \\ \hline
CheXNet\cite{rajpurkar2017chexnet} & 77.95 & 88.16 & 82.68 & 68.94 & 83.07 & 78.14 & 73.54 & 85.13 & 75.42 & 84.96 & 92.49 & 82.19 & 79.25 & 93.23 & 81.80 \\ \hline
DualCheXNet\cite{chen2019dualchexnet} & 78.40 & 88.80 & 83.10 & 70.50 & 83.80 & 79.60 & 72.70 & 87.60 & 74.60 & 85.20 & 94.20 & 83.70 & 79.60 & 91.20 & 82.36 \\ \hline
LLAGNet\cite{chen2019lesion} & 78.30 & 88.50 & 83.40 & 70.30 & 84.10 & 79.00 & 72.90 & 87.70 & 75.40 & 85.10 & 93.90 & 83.20 & 79.80 & 91.60 & 82.37 \\ \hline
Wang \etal\cite{wang2021triple} & 77.90 & 89.50 & 83.60 & 71.00 & 83.40 & 77.70 & 73.70 & 87.80 & 75.90 & 85.50 & 93.30 & 83.80 & 79.10 & \textcolor{red}{\textbf{93.80}} & 82.57 \\ \hline
Yan \etal\cite{yan2018weakly} & 79.24 & 88.14 & 84.15 & 70.95 & 84.70 & \textcolor{red}{\textbf{81.05}} & 73.97 & 87.59 & 75.98 & 84.70 & 94.22 & 83.26 & 80.83 & 93.41 & 83.01 \\ \hline
Luo \etal\cite{luo2020deep} & 78.91 & 90.69 & 84.18 & 71.84 & 83.76 & 79.85 & 74.19 & \textcolor{red}{\textbf{90.63}} & 76.81 & 86.10 & 93.96 & \textcolor{red}{\textbf{83.81}} & 80.36 & 93.71 & 83.49 \\ \hline
Arias-Garzon \etal\cite{arias2021covid} & 80.43 & 88.93 & 86.89 & 70.10 & 83.63 & 78.92 & 75.07 & 85.59 & 80.17 & 87.71 & 85.72 & 81.68 & 77.67 & 82.48 & 81.79 \\ \hline
Ouyang \etal\cite{ouyang2020learning}  & 77.00 & 87.00 & 83.00 & 71.00 & 83.00 & 79.00 & 72.00 & 88.00 & 74.00 & 84.00 & 94.00 & 83.00 & 79.00 & 91.00 & 81.79 \\ \hline
SDFN\cite{liu2019sdfn} & 78.10 & 88.50 & 83.20 & 70.00 & 81.50 & 76.50 & 71.90 & 86.60 & 74.30 & 84.20 & 92.10 & 83.50 & 79.10 & 91.10 & 81.47 \\ \hline
Keidar \etal\cite{keidar2021covid} & 80.64 & \textcolor{red}{\bf 90.88} & 86.94 & 70.60 & 83.93 & 77.07 & \textcolor{red}{\bf 76.53} & 85.54 & 80.43 & 89.20 & 90.87 & 81.47 & 78.02 & 91.80 & 83.14 \\ \hline
MANet\cite{xu2021manet} & 81.43 & 89.35 & 86.30 & 70.04 & 83.36 & 77.76 & 75.29 & 85.46 & 80.23 & 88.56 & 85.23 & 82.82 & 76.82 & 92.10 & 82.84 \\ \hline
PCAN \cite{Zhu2022PCAN} & 79.10 & 88.70 & 84.10 & 71.10 & 83.90 & 80.90 & 74.60 & 88.10 & 75.90 & 85.40 & \textcolor{red}{\bf 94.40} & 81.90 & 80.60 & 92.80 & 83.00\\ \hline
Proposed model & \textcolor{red}{\bf 82.68} & 90.16 & \textcolor{red}{\bf 88.35} & \textcolor{red}{\bf 72.34} & \textcolor{red}{\bf 86.73} & 80.70 & 76.38 & 88.98 & \textcolor{red}{\bf 81.16} & \textcolor{red}{\bf 90.78} & 92.70 & 82.56 & \textcolor{red}{\bf 81.29} & 90.53 & \textcolor{red}{\bf 84.67} \\ \hhline{================}
\end{tabularx}
\begin{tablenotes}
    \scriptsize 
    \item Here, Atel = Atelectasis, Card = Cardiomegaly, Effu = Effusion, Infi = Infiltration, Nodu = Nodule,  Pne1 = Pneumonia, Pne2 = Pneumothorax, Cons = Consolidation, Edem = Edema, Emph = Emphysema, Fibr = Fibrosis, PT = Pleural Thickening, Hern = Hernia
\end{tablenotes}
    \end{threeparttable}
\end{table*}

\begin{table*}[!t]
\caption{Comparison of disease classification AUC Scores (\%)  of the proposed model and SOTA models on the CheXpert dataset. The best results are shown in \textcolor{red}{red} font.}
\label{table:auc_chex}
\centering
\begin{tabular}{|c|c|c|c|c|c|c|}
\hline
    Model & Atelectasis & Cardiomegaly & Edema & Consolidation & Effusion & Mean \\ \hhline{=======}
    MANet \cite{xu2021manet} & 81.35 & 86.61 & 92.22 &91.59 & 89.86 & 88.33\\ \hline
    Arias-Garz´on \etal \cite{arias2021covid}&  81.74 & 84.24 & 94.06 & 90.74 & \textcolor{red}{\textbf{94.31}} & 89.02\\ \hline
    Keidar \etal \cite{keidar2021covid} & \textcolor{red}{\textbf{86.42}} & 87.39 & 91.97 & 88.23 & 91.73 & 89.15\\ \hline
    Irvin \etal \cite{irvin2019chexpert} & 85.80 & 83.20 & 94.10 & 89.90 & 93.40 & 89.30\\ \hline
    Pham \etal \cite{pham2021interpreting} & 82.50 & 85.50 & 93.00& 93.70 & 92.30 & 89.40\\ \hline
    ViT-S/16 \cite{Xiao2022DelvingIM} & 83.50 & 81.80 & 92.50 & \textcolor{red}{\textbf{94.50}} & 93.20 & 89.20 \\ \hline
    Zhu \etal \cite{Zhu2022PCAN} & 84.80 & 86.50 & 90.80 & 91.20 & 94.00 & 89.50 \\ \hline
    Proposed model & 86.21 & \textcolor{red}{\textbf{88.11}} &  \textcolor{red}{\textbf{94.15}} & 92.26 & 92.36 & \textcolor{red}{\textbf{90.62}} \\ \hline
\end{tabular}
\end{table*}

\begin{table*}[!t]
\caption{Ablation Study: Impact of different types of attention masks with respect to disease localization performance using different T(IoU) thresholds on the NIH dataset. The best results are shown in \textcolor{red}{red} font.}
\label{table: localization abl comp study}
\centering
\begin{threeparttable}
    \begin{tabular}{| c | c | c | c | c | c | c | c | c | c | c | c | c |}
    \hline
    T(IoU) & Method & AM (AbM) & AM (LM) & Atel & Card & Effu & Infil & Mass & Nodu & Pne1 & Pne2 & Mean \\
    \hline
    \multirow{4}{*}{  0.1} & Baseline &   &   & 0.6556 & \textcolor{red}{\textbf{1.0000}} & 0.8105 & 0.8293 & 0.7765 & 0.2911 & 0.7833 & 0.7732 & 0.7399 \\
     & ThoraX-PriorNet &   & \checkmark & 0.6889 & \textcolor{red}{\textbf{1.0000}} & 0.7974 & 0.8455 & 0.7176 & 0.4177 & 0.7917 & 0.7423 & 0.7504 \\
     & ThoraX-PriorNet & \checkmark &   & \textcolor{red}{\textbf{0.7333}} & \textcolor{red}{\textbf{1.0000}} & \textcolor{red}{\textbf{0.8366}} & 0.8293 & \textcolor{red}{\textbf{0.7882}} & \textcolor{red}{\textbf{0.5696}} & 0.8083 & \textcolor{red}{\textbf{0.8454}} & \textcolor{red}{\textbf{0.8013}} \\
     & ThoraX-PriorNet & \checkmark & \checkmark & \textcolor{red}{\textbf{0.7333}} & \textcolor{red}{\textbf{1.0000}} & 0.8235 & \textcolor{red}{\textbf{0.8780}} & 0.7294 & 0.4810 & \textcolor{red}{\textbf{0.8917}} & 0.7835 & 0.7901 \\ \hline
    \multirow{4}{*}{  0.2} & Baseline &   &   & 0.4333 & \textcolor{red}{\textbf{0.9726}} & 0.6209 & 0.6179 & \textcolor{red}{\textbf{0.6000}} & 0.1139 & 0.5750 & 0.5979 & 0.5664 \\
     & ThoraX-PriorNet &   & \checkmark & 0.4889 & 0.9041 & 0.6209 & 0.6829 & 0.5765 & 0.1772 & 0.6167 & 0.5052 & 0.5715 \\
     & ThoraX-PriorNet & \checkmark &   & \textcolor{red}{\textbf{0.5722}} & 0.8493 & \textcolor{red}{\textbf{0.7255}} & 0.5854 & 0.5882 & \textcolor{red}{\textbf{0.3038}} & \textcolor{red}{\textbf{0.7000}} & \textcolor{red}{\textbf{0.6701}} & 0.6243 \\
     & ThoraX-PriorNet & \checkmark & \checkmark & 0.5667 & 0.8973 & 0.6928 & \textcolor{red}{\textbf{0.7236}} & 0.5765 & 0.2532 & 0.6917 & 0.6082 & \textcolor{red}{\textbf{0.6262}} \\ \hline
    \multirow{4}{*}{  0.3} & Baseline &   &   & 0.2889 & \textcolor{red}{\textbf{0.7329}} & 0.4183 & 0.4553 & \textcolor{red}{\textbf{0.4706}} & 0.0380 & 0.4417 & 0.4330 & 0.4098 \\
     & ThoraX-PriorNet &   & \checkmark & 0.3444 & 0.6849 & 0.4248 & \textcolor{red}{\textbf{0.5447}} & 0.4235 & 0.0759 & 0.5000 & 0.3918 & 0.4238 \\
     & ThoraX-PriorNet & \checkmark &   & 0.4056 & 0.5342 & 0.5033 & 0.4715 & 0.4471 & \textcolor{red}{\textbf{0.1646}} & \textcolor{red}{\textbf{0.6083}} & \textcolor{red}{\textbf{0.5052}} & 0.4550 \\
     & ThoraX-PriorNet & \checkmark & \checkmark & \textcolor{red}{\textbf{0.4222}} & 0.5685 & \textcolor{red}{\textbf{0.5163}} & 0.5285 & 0.4588 & 0.1013 & 0.5667 & 0.4845 & \textcolor{red}{\textbf{0.4559}} \\ \hline
    \multirow{4}{*}{  0.4} & Baseline &   &   & 0.1667 & \textcolor{red}{\textbf{0.3425}} & 0.2418 & 0.3089 & 0.3529 & 0.0253 & 0.3083 & 0.3093 & 0.2570 \\
     & ThoraX-PriorNet &   & \checkmark & 0.2167 & 0.3288 & 0.2549 & 0.3333 & \textcolor{red}{\textbf{0.3882}} & 0.0127 & 0.3667 & 0.2784 & 0.2724 \\
     & ThoraX-PriorNet & \checkmark &   & \textcolor{red}{\textbf{0.3222}} & 0.2192 & 0.2876 & 0.3415 & 0.3529 & \textcolor{red}{\textbf{0.0886}} & \textcolor{red}{\textbf{0.4833}} & \textcolor{red}{\textbf{0.3814}} & 0.3096 \\
     & ThoraX-PriorNet & \checkmark & \checkmark & 0.2833 & 0.2603 & \textcolor{red}{\textbf{0.3464}} & \textcolor{red}{\textbf{0.4065}} & 0.3176 & 0.0380 & 0.4667 & \textcolor{red}{\textbf{0.3814}} & \textcolor{red}{\textbf{0.3125}} \\ \hline
    \multirow{4}{*}{  0.5} & Baseline &   &   & 0.0611 & 0.1233 & 0.0980 & 0.1870 & 0.2353 & 0.0127 & 0.1833 & 0.2062 & 0.1384 \\
     & ThoraX-PriorNet &   & \checkmark & 0.1111 & \textcolor{red}{\textbf{0.1438}} & 0.1307 & 0.2439 & 0.2235 & 0.0127 & 0.2917 & 0.1649 & 0.1653 \\
     & ThoraX-PriorNet & \checkmark &   & 0.1556 & 0.0959 & 0.1634 & 0.2358 & \textcolor{red}{\textbf{0.2706}} & \textcolor{red}{\textbf{0.0253}} & 0.2667 & 0.2268 & 0.1800 \\
     & ThoraX-PriorNet & \checkmark & \checkmark & \textcolor{red}{\textbf{0.1833}} & 0.1233 & \textcolor{red}{\textbf{0.2026}} & \textcolor{red}{\textbf{0.2602}} & 0.2353 & \textcolor{red}{\textbf{0.0253}} & \textcolor{red}{\textbf{0.3583}} & \textcolor{red}{\textbf{0.2784}} & \textcolor{red}{\textbf{0.2083}} \\ \hline
    \multirow{4}{*}{  0.6} & Baseline &   &   & 0.0278 & 0.0685 & 0.0260 & 0.1301 & 0.0588 & \textcolor{red}{\textbf{0.0127}} & 0.1333 & \textcolor{red}{\textbf{0.1340}} & 0.0739 \\
     & ThoraX-PriorNet &   & \checkmark & 0.0389 & \textcolor{red}{\textbf{0.0890}} & 0.0392 & 0.1301 & 0.0824 & 0.0000 & 0.2083 & 0.0515 & 0.0799 \\
     & ThoraX-PriorNet & \checkmark &   & \textcolor{red}{\textbf{0.0833}} & 0.0411 & 0.0588 & 0.0894 & \textcolor{red}{\textbf{0.1176}} & 0.0000 & 0.1500 & 0.1134 & 0.0817 \\
     & ThoraX-PriorNet & \checkmark & \checkmark & 0.0722 & 0.0411 & \textcolor{red}{\textbf{0.1111}} & \textcolor{red}{\textbf{0.1707}} & 0.0941 & 0.0000 & \textcolor{red}{\textbf{0.2667}} & 0.0973 & \textcolor{red}{\textbf{0.1061}} \\ \hline
    \multirow{4}{*}{   0.7} & Baseline &   &   & 0.0056 & \textcolor{red}{\textbf{0.0274}} & 0.0196 & 0.0325 & 0.0235 & 0.0000 & 0.0667 & 0.0412 & 0.0271 \\
     & ThoraX-PriorNet &   & \checkmark & 0.0111 & 0.0000 & 0.0065 & 0.0244 & \textcolor{red}{\textbf{0.0353}} & 0.0000 & \textcolor{red}{\textbf{0.1083}} & 0.0309 & 0.0271 \\
     & ThoraX-PriorNet & \checkmark &   & \textcolor{red}{\textbf{0.0167}} & 0.0068 & 0.0261 & 0.0407 & \textcolor{red}{\textbf{0.0353}} & 0.0000 & 0.0750 & \textcolor{red}{\textbf{0.0619}} & 0.0328 \\
     & ThoraX-PriorNet & \checkmark & \checkmark & 0.0222 & \textcolor{red}{\textbf{0.0274}} & \textcolor{red}{\textbf{0.0458}} & \textcolor{red}{\textbf{0.0813}} & 0.0235 & 0.0000 & 0.0125 & 0.0309 & \textcolor{red}{\textbf{0.0445}} \\
    \hline
    \end{tabular}
     \begin{tablenotes}
            \scriptsize 
            \item Here, AM (AbM) = APAM Utilizing Probabilistic Abnormality Mask,  AM (LM) = APAM Utilizing Chest ROI Mask, Atel = Atelectasis, Card = Cardiomegaly, Effu = Effusion, Infi = Infiltration, Nodu = Nodule,  Pne1 = Pneumonia, Pne2 = Pneumothorax
        \end{tablenotes}
	    \end{threeparttable}
\end{table*}

\begin{table*}[!t]
\caption{Ablation Study: Impact of input image spatial resolution with respect to disease localization performance using different T(IoU) thresholds on the NIH dataset. The best results are shown in \textcolor{red}{red} font.}
\label{table: localization abl study spatial resolution}
\centering
\begin{threeparttable}

\begin{tabular}{| c | c | c | c | c | c | c | c | c | c | c |}
\hline
T(IoU) & Method & Atel & Card & Effu & Infil & Mass & Nodu & Pne1 & Pne2 & Mean \\
\hline
\multirow{3}{*}{  0.1} & 224x224 & 0.6278 & \textcolor{red}{\textbf{1.0000}} & 0.7908 & 0.8618 & 0.5176 & 0.1266 & 0.8250 & 0.5773 & 0.6659 \\
 & 368x368 & 0.7111 & \textcolor{red}{\textbf{1.0000}} & 0.8170 & 0.8374 & 0.7059 & 0.2405 & 0.8000 & 0.7143 & 0.7283 \\
 & 512x512 & \textcolor{red}{\textbf{0.7611}} & \textcolor{red}{\textbf{1.0000}} & \textcolor{red}{\textbf{0.8366}} & \textcolor{red}{\textbf{0.8699}} & \textcolor{red}{\textbf{0.7412}} & \textcolor{red}{\textbf{0.5822}} & \textcolor{red}{\textbf{0.8667}} & \textcolor{red}{\textbf{0.7629}} & \textcolor{red}{\textbf{0.8026}} \\
 \hline
\multirow{3}{*}{  0.2} & 224x224 & 0.4611 & \textcolor{red}{\textbf{1.0000}} & 0.6209 & 0.6260 & 0.3176 & 0.0127 & \textcolor{red}{\textbf{0.6917}} & 0.3402 & 0.5088 \\
 & 368x368 & 0.5278 & 0.9932 & 0.6667 & 0.6748 & 0.5294 & 0.0380 & 0.6000 & 0.5816 & 0.5764 \\
 & 512x512 & \textcolor{red}{\textbf{0.5667}} & 0.8973 & \textcolor{red}{\textbf{0.6928}} & \textcolor{red}{\textbf{0.7236}} & \textcolor{red}{\textbf{0.5765}} & \textcolor{red}{\textbf{0.2532}} & \textcolor{red}{\textbf{0.6917}} & \textcolor{red}{\textbf{0.6082}} & \textcolor{red}{\textbf{0.6262}} \\
 \hline
\multirow{3}{*}{  0.3} & 224x224 & 0.3000 & \textcolor{red}{\textbf{0.9863}} & 0.4575 & 0.4390 & 0.2118 & 0.0000 & 0.5417 & 0.2474 & 0.3980 \\
 & 368x368 & 0.3500 & 0.8767 & 0.4706 & 0.5285 & 0.3765 & 0.0000 & 0.4583 & 0.3980 & 0.4323 \\
 & 512x512 & \textcolor{red}{\textbf{0.4667}} & 0.7945 & \textcolor{red}{\textbf{0.4902}} & \textcolor{red}{\textbf{0.4634}} & \textcolor{red}{\textbf{0.5059}} & \textcolor{red}{\textbf{0.1646}} & \textcolor{red}{\textbf{0.5583}} & \textcolor{red}{\textbf{0.4639}} & \textcolor{red}{\textbf{0.4884}} \\
 \hline
\multirow{3}{*}{  0.4} & 224x224 & 0.1722 & \textcolor{red}{\textbf{0.9452}} & 0.2614 & \textcolor{red}{\textbf{0.3089}} & 0.1647 & 0.0000 & 0.3917 & 0.1237 & 0.2960 \\
 & 368x368 & 0.2444 & 0.6304 & 0.2614 & 0.3821 & 0.2118 & 0.0000 & 0.2833 & 0.3163 & 0.2912 \\
 & 512x512 & \textcolor{red}{\textbf{0.3222}} & 0.6164 & \textcolor{red}{\textbf{0.2941}} & 0.2683 & \textcolor{red}{\textbf{0.3765}} & \textcolor{red}{\textbf{0.0506}} & \textcolor{red}{\textbf{0.4000}} & \textcolor{red}{\textbf{0.3402}} & \textcolor{red}{\textbf{0.3335}} \\
 \hline
\multirow{3}{*}{  0.5} & 224x224 & 0.1056 & \textcolor{red}{\textbf{0.7260}} & 0.1242 & \textcolor{red}{\textbf{0.2520}} & 0.1176 & 0.0000 & 0.2000 & 0.0825 & 0.2010 \\
 & 368x368 & 0.1111 & 0.2397 & 0.0980 & 0.2602 & 0.0706 & 0.0000 & 0.2000 & \textcolor{red}{\textbf{0.2347}} & 0.1518 \\
 & 512x512 & \textcolor{red}{\textbf{0.1722}} & 0.3904 & \textcolor{red}{\textbf{0.1438}} & 0.1789 & \textcolor{red}{\textbf{0.2941}} & \textcolor{red}{\textbf{0.0127}} & \textcolor{red}{\textbf{0.3167}} & 0.2268 & \textcolor{red}{\textbf{0.2169}} \\
 \hline
\multirow{3}{*}{  0.6} & 224x224 & 0.0556 & \textcolor{red}{\textbf{0.4658}} & 0.0719 & 0.1626 & 0.0235 & 0.0000 & 0.1083 & 0.0412 & \textcolor{red}{\textbf{0.1161}} \\
 & 368x368 & 0.0444 & 0.0890 & 0.0458 & 0.1382 & 0.0588 & 0.0000 & 0.1167 & \textcolor{red}{\textbf{0.1224}} & 0.0769 \\
 & 512x512 & \textcolor{red}{\textbf{0.0722}} & 0.0411 & \textcolor{red}{\textbf{0.1111}} & \textcolor{red}{\textbf{0.1707}} & \textcolor{red}{\textbf{0.0941}} & 0.0000 & \textcolor{red}{\textbf{0.2667}} & 0.0973 & 0.1061 \\
 \hline
\multirow{3}{*}{  0.7} & 224x224 & \textcolor{red}{\textbf{0.0222}} & \textcolor{red}{\textbf{0.1644}} & 0.0131 & 0.0650 & 0.0000 & 0.0000 & \textcolor{red}{\textbf{0.0583}} & 0.0103 & 0.0417 \\
 & 368x368 & 0.0111 & 0.0137 & 0.0196 & 0.0732 & \textcolor{red}{\textbf{0.0235}} & 0.0000 & 0.0250 & \textcolor{red}{\textbf{0.0510}} & 0.0271 \\
 & 512x512 & \textcolor{red}{\textbf{0.0222}} & 0.0274 & \textcolor{red}{\textbf{0.0458}} & \textcolor{red}{\textbf{0.0813}} & \textcolor{red}{\textbf{0.0235}} & 0.0000 & 0.0125 & 0.0309 & \textcolor{red}{\textbf{0.0445}} \\
\hline
\end{tabular}
     \begin{tablenotes}
            \item Here, Atel = Atelectasis, Card = Cardiomegaly, Effu = Effusion, Infi = Infiltration, Nodu = Nodule,  Pne1 = Pneumonia, Pne2 = Pneumothorax
        \end{tablenotes}

	    \end{threeparttable}
\end{table*}

\begin{table*}[!t]
\caption{Ablation Study: Effect of resizing feature and anatomy prior maps with respect to disease localization performance using different T(IoU) thresholds on the NIH dataset. The best results are shown in \textcolor{red}{red} font.}
\label{table: localization abl study feature map resizing}
\centering
\begin{threeparttable}
\begin{tabular}{| c | c | c | c | c | c | c | c | c | c | c |}
\hline
T(IoU) & Method & Atel & Card & Effu & Infil & Mass & Nodu & Pne1 & Pne2 & Mean \\
\hline
\multirow{3}{*}{ 0.1} & 16x16 & 0.7111 & 0.9932 & 0.8301 & \textcolor{red}{\textbf{0.8699}} & 0.7294 & 0.5190 & 0.8417 & 0.7523 & 0.7809 \\
 & 32x32 & \textcolor{red}{\textbf{0.7611}} & \textcolor{red}{\textbf{1.0000}} & \textcolor{red}{\textbf{0.8366}} & 0.8211 & 0.7176 & 0.4810 & 0.8583 & \textcolor{red}{\textbf{0.8247}} & 0.7876 \\
 & 48x48 & \textcolor{red}{\textbf{0.7611}} & \textcolor{red}{\textbf{1.0000}} & \textcolor{red}{\textbf{0.8366}} & \textcolor{red}{\textbf{0.8699}} & \textcolor{red}{\textbf{0.7412}} & \textcolor{red}{\textbf{0.5822}} & \textcolor{red}{\textbf{0.8667}} & 0.7629 & \textcolor{red}{\textbf{0.8026}} \\
 \hline
\multirow{3}{*}{ 0.2} & 16x16 & 0.5667 & 0.8767 & \textcolor{red}{\textbf{0.6928}} & \textcolor{red}{\textbf{0.6992}} & \textcolor{red}{\textbf{0.6235}} & 0.1899 & 0.6750 & 0.5979 & 0.6152 \\
 & 32x32 & \textcolor{red}{\textbf{0.6111}} & 0.9110 & 0.6405 & 0.6260 & 0.5529 & 0.2532 & \textcolor{red}{\textbf{0.7083}} & \textcolor{red}{\textbf{0.6495}} & 0.6191 \\
 & 48x48 & 0.5889 & \textcolor{red}{\textbf{0.9795}} & 0.6340 & 0.6341 & \textcolor{red}{\textbf{0.6235}} & \textcolor{red}{\textbf{0.3038}} & 0.6833 & 0.5567 & \textcolor{red}{\textbf{0.6255}} \\
 \hline
\multirow{3}{*}{ 0.3} & 16x16 & 0.4056 & 0.5890 & 0.4706 & \textcolor{red}{\textbf{0.5366}} & 0.4941 & 0.1013 & 0.5167 & 0.4536 & 0.4459 \\
 & 32x32 & 0.4389 & 0.5342 & 0.4837 & 0.4797 & 0.4353 & 0.1139 & 0.5500 & \textcolor{red}{\textbf{0.4845}} & 0.4400 \\
 & 48x48 & \textcolor{red}{\textbf{0.4667}} & \textcolor{red}{\textbf{0.7945}} & \textcolor{red}{\textbf{0.4902}} & 0.4634 & \textcolor{red}{\textbf{0.5059}} & \textcolor{red}{\textbf{0.1646}} & \textcolor{red}{\textbf{0.5583}} & 0.4639 & \textcolor{red}{\textbf{0.4884}} \\
 \hline
\multirow{3}{*}{ 0.4} & 16x16 & 0.2667 & 0.2877 & 0.2680 & \textcolor{red}{\textbf{0.3984}} & \textcolor{red}{\textbf{0.3765}} & 0.0380 & 0.3667 & \textcolor{red}{\textbf{0.4027}} & 0.3005 \\
 & 32x32 & 0.3000 & 0.1986 & \textcolor{red}{\textbf{0.3072}} & 0.2602 & 0.2588 & \textcolor{red}{\textbf{0.0633}} & 0.3750 & 0.3711 & 0.2668 \\
 & 48x48 & \textcolor{red}{\textbf{0.3222}} & \textcolor{red}{\textbf{0.6164}} & 0.2941 & 0.2683 & \textcolor{red}{\textbf{0.3765}} & 0.0506 & \textcolor{red}{\textbf{0.4000}} & 0.3402 & \textcolor{red}{\textbf{0.3335}} \\
 \hline
\multirow{3}{*}{ 0.5} & 16x16 & 0.1500 & 0.1644 & 0.1569 & \textcolor{red}{\textbf{0.2845}} & 0.2825 & \textcolor{red}{\textbf{0.0127}} & 0.2750 & 0.2165 & 0.1928 \\
 & 32x32 & \textcolor{red}{\textbf{0.1944}} & 0.0753 & \textcolor{red}{\textbf{0.1765}} & 0.1463 & 0.2000 & \textcolor{red}{\textbf{0.0127}} & 0.2083 & \textcolor{red}{\textbf{0.2680}} & 0.1602 \\
 & 48x48 & 0.1722 & \textcolor{red}{\textbf{0.3904}} & 0.1438 & 0.1789 & \textcolor{red}{\textbf{0.2941}} & \textcolor{red}{\textbf{0.0127}} & \textcolor{red}{\textbf{0.3167}} & 0.2268 & \textcolor{red}{\textbf{0.2169}} \\
 \hline
\multirow{3}{*}{ 0.6} & 16x16 & 0.0667 & 0.0616 & 0.0392 & \textcolor{red}{\textbf{0.1301}} & \textcolor{red}{\textbf{0.1529}} & 0.0000 & \textcolor{red}{\textbf{0.1917}} & \textcolor{red}{\textbf{0.1546}} & \textcolor{red}{\textbf{0.1000}} \\
 & 32x32 & 0.0611 & 0.0342 & \textcolor{red}{\textbf{0.0784}} & 0.0976 & 0.1294 & 0.0000 & 0.1417 & 0.1237 & 0.0833 \\
 & 48x48 & \textcolor{red}{\textbf{0.0778}} & \textcolor{red}{\textbf{0.1712}} & 0.0523 & 0.0894 & 0.1176 & 0.0000 & 0.1583 & 0.1031 & 0.0962 \\
 \hline
\multirow{3}{*}{ 0.7} & 16x16 & 0.0000 & 0.0205 & 0.0131 & \textcolor{red}{\textbf{0.0569}} & 0.0235 & 0.0000 & 0.0750 & 0.0616 & 0.0314 \\
 & 32x32 & \textcolor{red}{\textbf{0.0056}} & 0.0068 & \textcolor{red}{\textbf{0.0458}} & 0.0325 & 0.0235 & 0.0000 & 0.0417 & \textcolor{red}{\textbf{0.0619}} & 0.0272 \\
 & 48x48 & 0.0000 & \textcolor{red}{\textbf{0.0753}} & 0.0131 & 0.0244 & \textcolor{red}{\textbf{0.0352}} & 0.0000 & \textcolor{red}{\textbf{0.0833}} & 0.0412 & \textcolor{red}{\textbf{0.0341}} \\
\hline
\end{tabular}
     \begin{tablenotes}
            \item Here, Atel = Atelectasis, Card = Cardiomegaly, Effu = Effusion, Infi = Infiltration, Nodu = Nodule,  Pne1 = Pneumonia, Pne2 = Pneumothorax
        \end{tablenotes}
	    \end{threeparttable}
\end{table*}

\begin{table*}[!t]
\caption{Comparison of disease localization accuracy of the best performing proposed model with state-of-the-art methods. The best results are shown in \textcolor{red}{red} font.}
\label{table:sota comp localization NIH}
\centering
\begin{threeparttable}
\begin{tabular}{| c | c | c | c | c | c | c | c | c | c | c | c |}
\hline
T(IoU)  & Method & Atel & Card & Effu & Infil & Mass & Nodu & Pne1 & Pne2 & Mean \\
\hline
\multirow{10}{*}{   0.1} & Cai \etal \cite{cai2018iterative} & 0.68 & 0.97 & 0.65 & 0.52 & 0.56 & 0.46 & 0.65 & 0.43 & 0.61 \\ 
& Li \etal \cite{Li2017ThoracicSupervision} & 0.59 & 0.81 & 0.72 & 0.84 & 0.68 & 0.28 & 0.22 & 0.37 & 0.57 \\
& Liu \etal \cite{liu2019align} & 0.39 & 0.90 & 0.63 & 0.85 & 0.69 & 0.38 & 0.30 & 0.39 & 0.60 \\
& Ouyang \etal \cite{Ouyang2021LearningDiagnosis} & 0.78 & 0.97 & 0.82 & 0.85 & 0.78 & 0.56 & 0.76 & 0.48 & 0.75 \\
& Han \etal \cite{yan2021Cross} \dag & 0.72 & 0.96 & 0.88 & 0.93 & 0.74 & 0.45 & 0.65 & 0.64 & 0.75 \\
& Han \etal \cite{9930800} \ddag & 0.61 & 0.95 & 0.65 & 0.82 & 0.50 & 0.13 & 0.79 & 0.28 & 0.59\\
 & Li \etal \cite{Li2022Model} & 0.64 & \textcolor{red}{\textbf{1.00}} & 0.75 & 0.79 & 0.69 & 0.07 & 0.79 & 0.39 & 0.64 \\
 & Zhu \etal \cite{Zhu2022PCAN} & \textcolor{red}{\textbf{0.84}} & \textcolor{red}{\textbf{1.00}} & \textcolor{red}{\textbf{0.86}} & \textcolor{red}{\textbf{0.94}} & \textcolor{red}{\textbf{0.82}} & \textcolor{red}{\textbf{0.49}} & \textcolor{red}{\textbf{0.90}} & 0.38 & 0.78 \\
& Rozenberg \etal \cite{Rozenberg2021} \dag & 0.77 & \textcolor{red}{\textbf{1.00}} & 0.84 & 0.94 & 0.70 & 0.44 & 0.91 & 0.73 & 0.79 \\ 
& Proposed model & 0.73 & \textcolor{red}{\textbf{1.00}} & 0.82 & 0.88 & 0.73 & 0.48 & 0.89 & \textcolor{red}{\textbf{0.78}} & \textcolor{red}{\textbf{0.80}} \\ \hline
\multirow{6}{*}{   0.2}& Cai \etal \cite{cai2018iterative} & 0.51 & 0.90 & 0.52 & 0.44 & 0.47 & 0.27 & 0.54 & 0.24 & 0.49 \\
& Han \etal \cite{yan2021Cross} \dag & 0.55 & 0.89 & 0.78 & 0.85 & 0.62 & 0.31 & 0.52 & 0.54 & 0.63 \\
& Han \etal \cite{9930800} \ddag & 0.41 & 0.91 & 0.41 & 0.59 & 0.26 & 0.05 & 0.57 & 0.19 & 0.42\\
& Li \etal \cite{Li2022Model} & 0.40 & \textcolor{red}{\textbf{1.00}} & 0.66 & \textcolor{red}{\textbf{0.74}} & 0.43 & 0.01 & \textcolor{red}{\textbf{0.69}} & 0.28 & 0.53 \\
& Zhu \etal \cite{Zhu2022PCAN} & 0.47 & 0.68 & 0.45 & 0.48 & 0.26 & 0.05 & 0.35 & 0.23 & 0.37 \\
& Proposed model & \textcolor{red}{\textbf{0.57}} & 0.90 & \textcolor{red}{\textbf{0.69}} & 0.72 & \textcolor{red}{\textbf{0.58}} & \textcolor{red}{\textbf{0.25}} & \textcolor{red}{\textbf{0.69}} & \textcolor{red}{\textbf{0.61}} & \textcolor{red}{\textbf{0.63}} \\
\hline
\multirow{9}{*}{        0.3} & Cai \etal \cite{cai2018iterative} & 0.33 & 0.85 & 0.34 & 0.28 & 0.33 & 0.11 & 0.39 & 0.16 & 0.35 \\
& Li \etal \cite{Li2017ThoracicSupervision} & 0.34 & 0.26 & \textcolor{red}{\textbf{0.52}} & \textcolor{red}{\textbf{0.72}} & 0.40 & 0.09 & 0.00 & 0.23 & 0.32 \\
& Liu \etal \cite{liu2019align} & 0.34 & 0.71 & 0.39 & 0.65 & 0.48 & 0.09 & 0.16 & 0.20 & 0.38 \\
& Ouyang \etal \cite{Ouyang2021LearningDiagnosis} & 0.34 & 0.40 & 0.27 & 0.55 & \textcolor{red}{\textbf{0.51}} & \textcolor{red}{\textbf{0.14}} & 0.42 & 0.22 & 0.36 \\
& Han \etal \cite{yan2021Cross} \dag & 0.39 & 0.85 & 0.60 & 0.67 & 0.43 & 0.21 & 0.40 & 0.45 & 0.50 \\
& Han \etal \cite{9930800} \ddag & 0.28 & 0.79 & 0.22 & 0.38 & 0.12 & 0.01 & 0.41 & 0.05 & 0.28\\
& Li \etal \cite{Li2022Model} & 0.21 & \textcolor{red}{\textbf{1.00}} & 0.44 & 0.53 & 0.27 & 0.00 & 0.55 & 0.19 & 0.40 \\
& Zhu \etal \cite{Zhu2022PCAN} & \textcolor{red}{\textbf{0.43}} & 0.34 & 0.33 & 0.57 & 0.48 & 0.04 & \textcolor{red}{\textbf{0.60}} & 0.13 & 0.36 \\
& Proposed model & 0.42 & 0.57 & \textcolor{red}{\textbf{0.52}} & 0.53 & 0.46 & 0.10 & 0.57 & \textcolor{red}{\textbf{0.48}} & \textcolor{red}{\textbf{0.49}} \\
\hline
\multirow{6}{*}{        0.4}& Cai \etal \cite{cai2018iterative} & 0.23 & 0.73 & 0.18 & 0.20 & 0.18 & 0.03 & 0.23 & 0.11 & 0.24 \\
& Han \etal \cite{yan2021Cross} \dag & 0.24 & 0.81 & 0.42 & 0.54 & 0.34 & 0.13 & 0.28 & 0.32 & 0.39 \\
& Han \etal \cite{9930800} \ddag & 0.17 & 0.54 & 0.13 & 0.18 & 0.07 & 0.01 & 0.26 & 0.02 & 0.17\\
& Li \etal \cite{Li2022Model} & 0.10 & \textcolor{red}{\textbf{0.98}} & 0.27 & \textcolor{red}{\textbf{0.47}} & 0.18 & 0.00 & 0.38 & 0.13 & 0.31 \\
& Zhu \etal \cite{Zhu2022PCAN} & 0.23 & 0.10 & 0.16 & 0.37 & 0.37 & 0.01 & 0.33 & 0.10 & 0.21 \\
& Proposed model & \textcolor{red}{\textbf{0.32}} & 0.62 & \textcolor{red}{\textbf{0.29}} & 0.27 & \textcolor{red}{\textbf{0.38}} & \textcolor{red}{\textbf{0.05}} & \textcolor{red}{\textbf{0.40}} & \textcolor{red}{\textbf{0.34}} & \textcolor{red}{\textbf{0.33}} \\
\hline
\multirow{8}{*}{    0.5}& Cai \etal \cite{cai2018iterative} & 0.11 & 0.60 & 0.10 & 0.12 & 0.07 & \textcolor{red}{\textbf{0.03}} & 0.17 & 0.08 & 0.17 \\
& Li \etal \cite{Li2017ThoracicSupervision} & 0.18 & 0.10 & \textcolor{red}{\textbf{0.27}} & 0.46 & 0.18 & \textcolor{red}{\textbf{0.03}} & 0.00 & 0.11 & 0.17 \\
& Liu \etal \cite{liu2019align} & \textcolor{red}{\textbf{0.19}} & 0.53 & 0.19 & \textcolor{red}{\textbf{0.47}} & \textcolor{red}{\textbf{0.33}} & \textcolor{red}{\textbf{0.03}} & 0.08 & 0.11 & \textcolor{red}{\textbf{0.24}} \\
& Han \etal \cite{yan2021Cross} \dag & 0.16 & 0.77 & 0.29 & 0.35 & 0.24 & 0.09 & 0.15 & 0.22 & 0.28 \\
& Han \etal \cite{9930800} \ddag & 0.08 & 0.32 & 0.05 & 0.09 & 0.05 & 0.00 & 0.12 & 0.01 & 0.09\\
& Li \etal \cite{Li2022Model} & 0.05 & \textcolor{red}{\textbf{0.87}} & 0.13 & 0.34 & 0.12 & 0.00 & 0.33 & 0.10 & 0.24 \\
& Zhu \etal \cite{Zhu2022PCAN} & 0.09 & 0.01 & 0.11 & 0.19 & 0.21 & 0.00 & 0.17 & 0.05 & 0.10 \\
& Proposed model & 0.18 & 0.12 & 0.20 & 0.26 & 0.24 & \textcolor{red}{\textbf{0.03}} & \textcolor{red}{\textbf{0.36}} & \textcolor{red}{\textbf{0.28}} & 0.22 \\
\hline
\multirow{5}{*}{  0.6}& Cai \etal \cite{cai2018iterative} & 0.03 & 0.44 & 0.05 & 0.06 & 0.05 & \textcolor{red}{\textbf{0.01}} & 0.05 & 0.07 & 0.10 \\
& Han \etal \cite{yan2021Cross} \dag & 0.09 & 0.74 & 0.19 & 0.16 & 0.18 & 0.04 & 0.11 & 0.14 & 0.21\\
& Han \etal \cite{9930800} \ddag & 0.02 & 0.15 & 0.03 & 0.04 & 0.03 & 0.00 & 0.06 & 0.00 & 0.04\\
& Li \etal \cite{Li2022Model} & 0.01 & \textcolor{red}{\textbf{0.60}} & 0.06 & \textcolor{red}{\textbf{0.23}} & 0.06 & 0.00 & 0.17 & 0.04 & \textcolor{red}{\textbf{0.15}} \\
& Proposed model & \textcolor{red}{\textbf{0.07}} & 0.04 & \textcolor{red}{\textbf{0.11}} & 0.17 & \textcolor{red}{\textbf{0.09}} & 0.00 & \textcolor{red}{\textbf{0.27}} & \textcolor{red}{\textbf{0.10}} & 0.11 \\
\hline
\multirow{7}{*}{    0.7} & Cai \etal \cite{cai2018iterative} & 0.01 & 0.17 & 0.01 & 0.02 & 0.01 & 0.00 & 0.02 & 0.02 & 0.03 \\
& Li \etal \cite{Li2017ThoracicSupervision} & \textcolor{red}{\textbf{0.09}} & 0.01 & 0.07 & \textcolor{red}{\textbf{0.28}} & 0.08 & \textcolor{red}{\textbf{0.01}} & 0.00 & 0.05 & 0.07 \\
& Liu \etal \cite{liu2019align} & 0.08 & \textcolor{red}{\textbf{0.30}} & \textcolor{red}{\textbf{0.09}} & 0.25 & \textcolor{red}{\textbf{0.19}} & \textcolor{red}{\textbf{0.01}} & 0.04 & \textcolor{red}{\textbf{0.07}} & \textcolor{red}{\textbf{0.13}} \\
 & Han \etal \cite{yan2021Cross} \dag & 0.05 & 0.54 & 0.09 & 0.11 & 0.12 & 0.02 & 0.07 & 0.06 & 0.13 \\
& Han \etal \cite{9930800} \ddag & 0.01 & 0.04 & 0.01 & 0.02 & 0.01 & 0.00 & 0.03 & 0.00 & 0.02\\
& Li \etal \cite{Li2022Model} & 0.01 & 0.26 & 0.02 & 0.10 & 0.02 & 0.00 & \textcolor{red}{\textbf{0.10}} & 0.02 & 0.07 \\
& Proposed model & 0.02 & 0.03 & 0.05 & 0.08 & 0.02 & 0.00 & 0.01 & 0.03 & 0.04 \\
\hline
\end{tabular}
\begin{tablenotes}
            \item {\dag} utilized bounding box information {\ddag} scores for input image resolution of 224x224 
            \item Here, Atel = Atelectasis, Card = Cardiomegaly, Effu = Effusion, Infi = Infiltration, Nodu = Nodule,  Pne1 = Pneumonia, Pne2 = Pneumothorax
        \end{tablenotes}
	    \end{threeparttable}
\end{table*}
\copyrightnotice
\section{Experimental Results}
\label{sec:Experimental Results}
\subsection{Disease Classification}
\subsubsection{Ablation Study} 
We have conducted several ablation studies on the NIH ChestX-ray14 dataset of our trained model for different thoracic abnormalities. First, we evaluate the impact of attention masks, i.e., probabilistic abnormality mask and chest ROI mask, on the classification performance. Table \ref{table: classification abl comp} shows the reported results. The baseline model showed a mean AUC (\%) score of 84.30, which performed better for classifying diseases like Atelectasis, Nodules, Consolidation, and Edema. The baseline denotes the vanilla DenseNet121 model without incorporating the APAM block. Afterward, we added the APAM block and gradually used the different types of attention masks. Table \ref{table: classification abl comp} demonstrates that all three ThoraX-PriorNet variants achieve better classification scores than the baseline. We obtained the most significant jump in classification results when we used APAM with the probabilistic disease-specific masks, i.e., an AUC (\%) score of 84.69. Incorporating both types of attention masks yields a slightly lower score, i.e., a percentage AUC score of 84.67. But it improves the performance for pathologies like Effusion, Infiltration, Fibrosis, and Pleural Thickening.\par
\copyrightnotice
Next, we conducted an ablation study to explore the impact of the input image sizes on the classification performance. We resize the input image into three different sizes: 256\texttimes256, 420\texttimes420, and 586\texttimes586 and crop 224\texttimes224 patches for 256\texttimes256, 368\texttimes368 patches for 420\texttimes420, and 512\texttimes512 patches for 586\texttimes586 as inputs, respectively (random crop during training, center crop during inference). Results on the NIH Chest X-ray dataset are shown in Table \ref{table: classification abl spatial resolution}. We can see that increasing input image resolution improves the classification performance. However, the improvement range from 368\texttimes368 to 512\texttimes512 is lower compared to 224\texttimes224 to 368x368. More specifically, we observe that the increase in AUC score for small lesions, such as nodules, is significant in the higher resolution.\par
Finally, we conducted an ablation study on the spatial dimension of the feature map and the attention masks. We downscale and upsample the attention masks and feature maps, respectively, to an intermediate size before using them in the APAM block. For the input image dimension of 512\texttimes512, the feature map size is 16\texttimes16. We also performed experiments by resizing the feature map to 32\texttimes32 and 48\texttimes48. The results are reported in Table \ref{table: classification abl feature map resizing}. We observe that increasing the spatial dimension of the final feature map does not yield improvements in the classification performance. The 48\texttimes48 model has the same classification performance level as the 16\texttimes16 model. However, the 48\texttimes48 model improves the localization performance, which will be demonstrated in a later section.
\begin{figure*}[!t]
    \centering
    \includegraphics[width=0.9\textwidth]{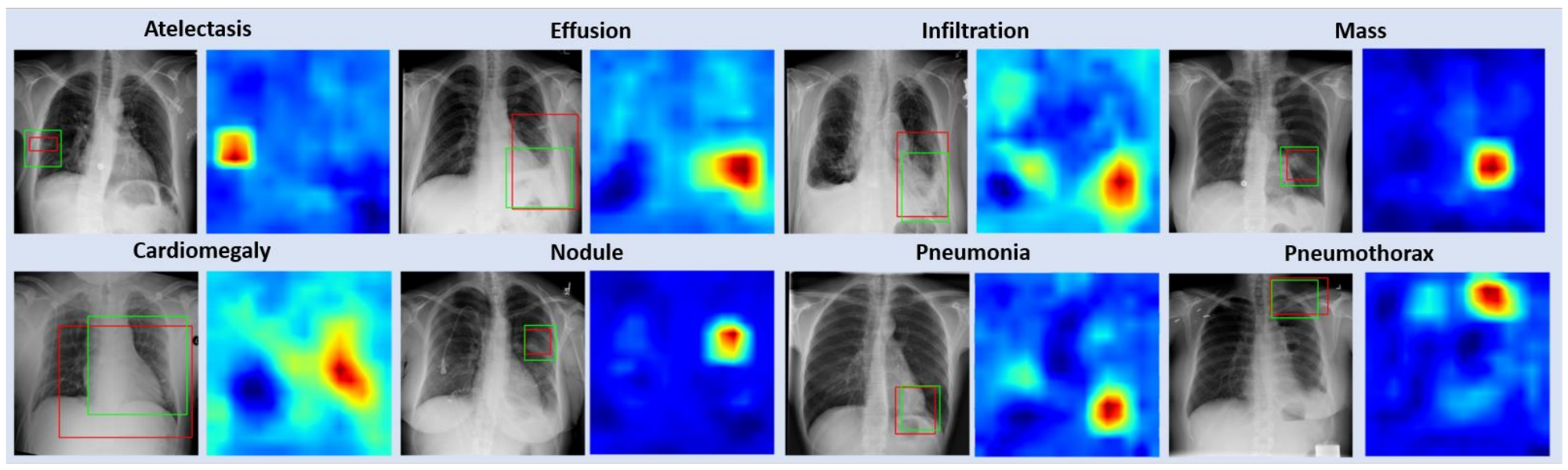}
    \caption{Examples of some disease localization by our proposed method. The first column of each sample: Input CXR image with the ground truth bounding box (red color) and the predicted bounding box (green color). The second column of each sample: Corresponding activation map from the proposed model.}
    \label{fig:heatmap}
\end{figure*}

\subsubsection{Performance Comparison with SOTA Methods}
Table \ref{table:auc_compare} compares the AUC score of ThoraX-PriorNet with other state-of-the-art (SOTA) models on NIH ChestX-ray14 dataset. Here, we observe that the proposed model's performance is superior to existing SOTA methods in terms of the mean AUC score. More specifically, it has shown performance improvement in diseases like Atelectasis, Effusion, Infiltration, Mass, Consolidation, Edema, and Pleural Thickening. \par
Table \ref{table:auc_chex} shows the comparison of our proposed model with existing state of the art models on CheXpert dataset. Here, we have used the same probabilistic masks which were generated for training on NIH Chest X-ray dataset and for providing disease guided attention. The results show that the proposed method provides superior results for diseases like- cardiomegaly and edema, whereas performance on atelectasis, consolidation, and effusion are slightly less than the compared approaches. However, the overall mean AUC score is better compared to the other models. Our method shows an AUC score of 90.62\%.
\copyrightnotice
\subsection{Abnormality Localization}
\subsubsection{Ablation Study} 
We have also conducted several ablation studies on the NIH ChestX-ray14 dataset to explore the impact of different aspects of our trained model on localization performance. First, we evaluate the impact of different types of attention masks. The results are reported in Table~\ref{table: localization abl comp study}. We can observe a notable performance improvement after including the APAM module. Our proposed ThoraX-PriorNet outperforms the baseline model by large margins in all T(IoU) thresholds. APAM block utilizing both disease probabilistic maps and chest ROI maps achieves overall better results, especially in the higher thresholds compared to the APAM block using only one type of attention mask.\par
The impact of different input image resolutions on the localization performance is demonstrated in Table~\ref{table: localization abl study spatial resolution}. We can observe that increasing the spatial dimension of the input image enhances the localization performance greatly. More specifically, we observe that increasing spatial dimension shows greater performance improvement in the localization tasks for diseases with small spatial features (e.g., mass, nodule, pneumothorax). However, large lesions, such as cardiomegaly, are not benefited. The impact on localization performance due to different dimensions of the intermediate size of feature maps and attention maps is reported in Table~\ref{table: localization abl study feature map resizing}. Similar to the input spatial dimension, we can observe that the 48x48 model achieved overall better localization performance compared to other models.
\begin{table*}[!t]
\caption{Statistical analysis between the baseline and proposed model for a 10-fold cross-validation using Nadeau and Bengio’s corrected t-test Method \cite{Nadeau2003}.}
\label{table: statistical significance test}
\centering
\begin{adjustbox}{width=\textwidth}
\begin{tabular}{| c | c | c | c | c | c | c | c | c | c | c | c | c |}
\hline
Method & Test-1 & Test-2 & Test-3 & Test-4 & Test-5 & Test-6 & Test-7 & Test-8 & Test-9 & Test-10 & Mean $\pm$ Std & p-value\\
\hline
Baseline & 84.56 & 84.38 & 84.13 & 84.81 & 84.21 & \textcolor{red}{\textbf{84.09}} & 84.79 & 84.12 & 84.65 & 84.33 & 84.41 $\pm$ 0.26 & \multirow{2}{*}{0.017}\\
\cline{1-12}
Proposed & \textcolor{red}{\textbf{84.66}} & \textcolor{red}{\textbf{84.60}} & \textcolor{red}{\textbf{84.51}} & \textcolor{red}{\textbf{84.97}} & \textcolor{red}{\textbf{84.65}} & 84.03 & \textcolor{red}{\textbf{84.99}} & \textcolor{red}{\textbf{84.34}} & \textcolor{red}{\textbf{84.82}} & \textcolor{red}{\textbf{84.58}} & \textcolor{red}{\textbf{84.61 $\pm$ 0.27}} & \\
\hline
\end{tabular}    
\end{adjustbox}
\end{table*}
\copyrightnotice
\subsubsection{Performance Comparison with SOTA Methods} 
Table \ref{table:sota comp localization NIH} shows the quantitative comparison of the localization score of ThoraX-PriorNet with previous SOTA models. Note that Han \etal and Rozenberg \etal utilize bounding box information in their pipeline. As a result, their model is not directly comparable to ours and other SOAT models. In spite of that, our proposed method shows comparable performance at lower T(IoU) thresholds despite not using the bounding box supervision. Our proposed ThoraX-PriorNet achieved improvements of 2.56\%, 18.87\%, 22.50\%, and 6.45\% at IoU of 0.1, 0.2, 0.3, and 0.4, respectively, compared to the localization performances of existing methods. In other IoU thresholds, our model achieves slightly lower but competitive scores.\par
We have extracted the activation maps for eight different diseases from the NIH ChestX-ray8 dataset and plotted them in Fig. \ref{fig:heatmap} to visualize the localization of the proposed model. The red boxes denote the ground truth boxes, while the green boxes denote the predicted boxes. We can observe that our model can identify and localize the abnormal findings.

\subsection{Statistical Analysis}
To perform statistical analysis, we have conducted a 10-fold cross-validation and used Nadeau and Bengio’s corrected t-test method \cite{Nadeau2003} for calculating the p-values. The results for the baseline and the proposed method are reported in Table \ref{table: statistical significance test}. The baseline model achieves an average AUC (\%) score of 84.41 with a standard deviation of 0.26, while our proposed method achieves 84.61\textpm0.27. The statistical result yields a p-value of 0.017, denoting the improvement of the proposed method compared to the baseline.

\subsection{COMPUTATIONAL COMPLEXITY ANALYSIS}
The average time to process a single chest X-ray image during the testing phase, along with the floating point operation computation, for the input image dimension of 512\texttimes512 is reported in Table \ref{table: comp cost}. Our proposed  ThoraX-PriorNet takes an average of 8.74 ms to process a test image and requires 28.1 GFLOPS compute power to perform this task.
\begin{table}[!t]
\caption{Computational cost parameters by ThoraX-PriorNet for a single image on 512x512 dimension during the test phase on the NIH ChestX-Ray14 dataset.}
\label{table: comp cost}
\centering
\begin{tabular}{| c | c | c | c |}
\hline
Method & Time (ms) & FLOPs (G) & AUC\\
\hline
ThoraX-PriorNet & 8.74 & 28.1 & 84.67\\
\hline
\end{tabular}    
\end{table}

\begin{figure*}[!t]
    \centering
    \includegraphics[width=0.9\linewidth]{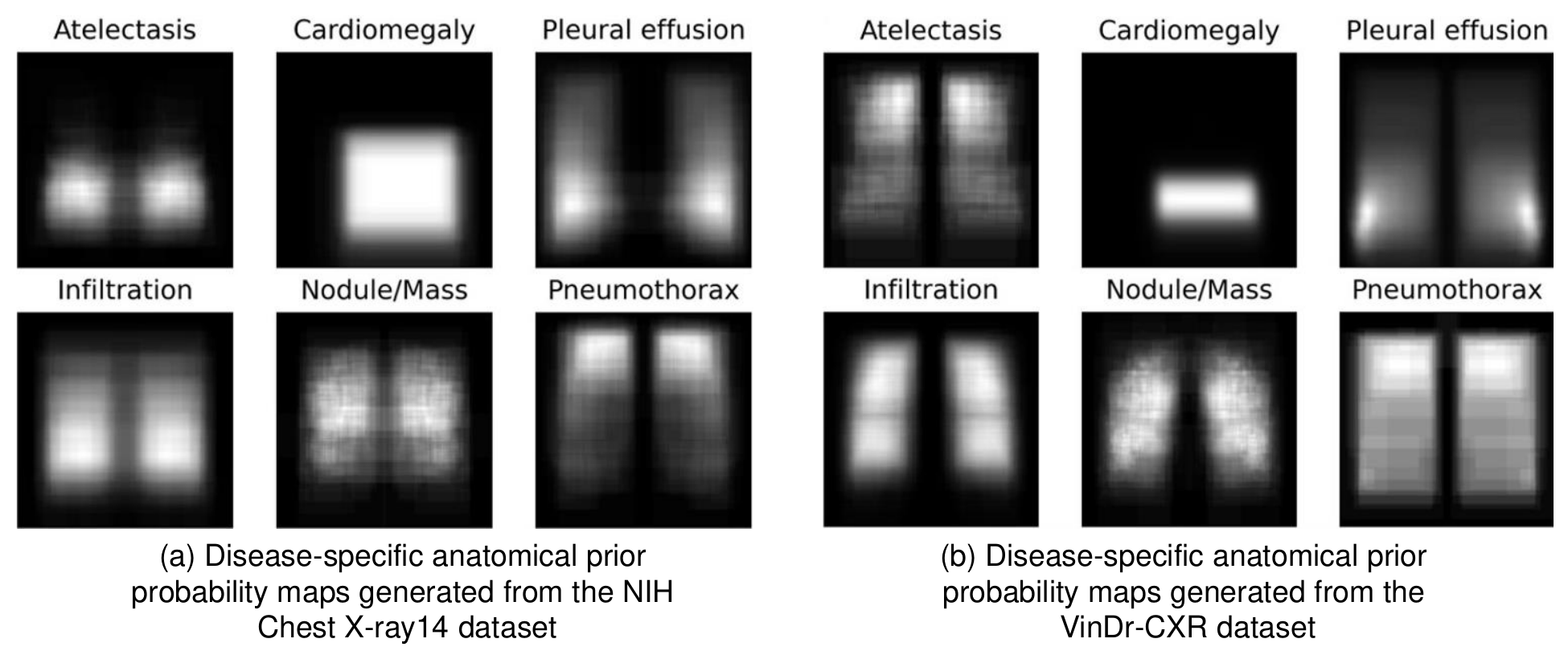}
    \caption{Disease-specific anatomical prior probability maps. Abnormality masks generated from NIH (left) and VinDr-CXR (right) datasets.}
    \label{fig:abmask generalizability test}
\end{figure*}

\begin{table*}[!t]
\caption{Evaluation of the generalizability of the disease-specific anatomical prior probability maps across different thoracic disease datasets.}
\label{table:abmask generalizability test}
\centering
\begin{adjustbox}{width=\textwidth}
\begin{threeparttable}
\begin{tabular}{| c  | c  | c  | c  |c  |c  |c  |c  |c  |c  |c |c |}
\hline
  &   & \multicolumn{2}{c}{Abnormality masks used} & \multicolumn{7}{|c|}{T(IoU)} & \multirow{2}{*}{ \makecell{Average of \\ per threshold \%RI}} \\ \cline{3-11}
Dataset & Method & NIH CXR14 & VinDr-CXR &  0.1 &  0.2 &  0.3 &  0.4 &  0.5 &  0.6 &  0.7 & \\ 
\hline
\multirow{4}{*}{NIH CXR14} & DenseNet-121 &   &   & 0.4387 & 0.3077 & 0.2036 & 0.1170 & 0.0671 & 0.0359 & 0.0082 & REF \\ 
\cline{2-12} 
 & DenseNet-121+Align &   &   & 0.4665 & 0.3552 & 0.2457 & 0.1540 & 0.0676 & 0.0178 & 0.0064 & +0.35\% \\ 
\cline{2-12}  
 & ThoraX-PriorNet & \checkmark &   & \textcolor{red}{\textbf{0.6897}} & \textcolor{red}{\textbf{0.5446}} & \textcolor{red}{\textbf{0.3921}} & \textcolor{red}{\textbf{0.2226}} & \textcolor{red}{\textbf{0.1212}} & \textcolor{red}{\textbf{0.0487}} & 0.0074 & +60.51\% \\ 
\cline{2-12}  
 & ThoraX-PriorNet &   & \checkmark & 0.6798 & 0.5336 & 0.3666 & 0.2028 & 0.0943 & 0.0476 & \textcolor{red}{\textbf{0.0139}} & \textcolor{red}{\textbf{+60.63\%}} \\ 
\hline
\multirow{4}{*}{VinDr-CXR} & DenseNet-121 &   &   & 0.4008 & 0.2752 & 0.1702 & 0.0646 & 0.0275 & 0.0209 & 0.0060 & REF \\
\cline{2-12} 
 & DenseNet-121+Align &   &   & 0.4016 & 0.2933 & 0.2019 & 0.1043 & 0.0432 & 0.0136 & 0.0000 & +1.29\% \\ 
\cline{2-12}  
 & ThoraX-PriorNet & \checkmark &   & 0.4536 & 0.3452 & 0.2325 & 0.1262 & 0.0413 & 0.0192 & 0.0006 & +17.52\% \\
\cline{2-12}  
 & ThoraX-PriorNet &   & \checkmark & \textcolor{red}{\textbf{0.4565}} & \textcolor{red}{\textbf{0.3536}} & \textcolor{red}{\textbf{0.2468}} & \textcolor{red}{\textbf{0.1446}} & \textcolor{red}{\textbf{0.0598}} & \textcolor{red}{\textbf{0.0198}} & \textcolor{red}{\textbf{0.0033}} & \textcolor{red}{\textbf{+39.77\%}} \\ 
 \hline 
\end{tabular}
\begin{tablenotes}
            \item Here, \%RI = \%Relative improvement
        \end{tablenotes}
	    \end{threeparttable}
     \end{adjustbox}
\end{table*}
\copyrightnotice
\subsection{Analysis of Generaziability of the Probabilistic Abnormality Masks}
Different chest X-ray-based thoracic disease datasets may have diverse affine variations, such as rotations, shifts, and different scales. To address the affine variations, we have utilized the alignment module \cite{liu2019align}. In addition, the chest X-ray datasets may have intrinsic variations among them due to patient demographics, geographical diversity, class imbalances, different exposure settings and imaging protocols, scanner intrinsic variations, and so on, inherent to medical datasets. However, in our experiments, we are not utilizing or training different datasets together, a task that is reserved for domain adaptation and generalization methods \cite{WANG2023104488, 9557808, zunaed2023learning}. Here, we are generating the disease-prior masks by taking and aggregating the referenced spatial positions from the bounding boxes to get a probabilistic map. The domain variations due to exposure shift, different imaging protocols, or machine intrinsic variations are not propagated through the generated abnormality masks. However, we do acknowledge that the number and quality of the ground truth bounding boxes, class imbalance, patient demographics, or geographical diversity may have an effect on the generated probabilistic map, which may influence the performance of the proposed model.\par
We have conducted experiments to evaluate the generalizability of the disease-prior probabilistic abnormality masks generated from a particular thoracic disease dataset. For this experiment, we have chosen the NIH chest X-ray14 \cite{wang2017chestx} and the VinDr-CXR dataset \cite{nguyen2020vindrcxr}, as they have provided bounding box annotations. We have performed the experiment for the six common pathologies between them, i.e., Atelectasis, Cardiomegaly, Pleural Effusion, Infiltration, Nodule/Mass, and Pneumothorax. For the NIH chest X-ray14 dataset, we have merged the Nodule and Mass classes into a single Nodule/Mass class, similar to the VinDr-CXR dataset. The VinDr-CXR dataset has a much higher number of available ground truth bounding boxes compared to the NIH chest X-ray14. The generated disease prior masks from NIH chest X-ray14 and VinDr-CXR are given in Fig.~\ref{fig:abmask generalizability test}. Note that we have utilized only the bounding boxes from the official training split of the VinDr-CXR to generate the disease masks.\par
First, we train the vanilla DenseNet-121 model on both datasets without using the aligned images. Afterward, we train the vanilla DenseNet-121 with the aligned images. Finally, we train our proposed ThoraX-PriorNet, utilizing the dataset-specific abnormality masks from the NIH chest X-ray14 and VinDr-CXR datasets, one at a time. The results are reported in Table \ref{table:abmask generalizability test}. The average improvement is calculated as follows:
\begin{equation}
    (\text{\%RI})_{\text{avg}} = \frac{1}{n}\sum_{i=1}^{n}\frac{S_i-S_i^{\text{ref}}}{S_i^{\text{ref}}}*100 
\end{equation}
Here, $n$ is the number of thresholds, $S_i$ is the performance at a particular threshold $i$, and $S_i^{\text{ref}}$ is the performance of the vanilla DenseNet-121 at threshold $i$. We can observe that adding the alignment module improves the performance of the vanilla DenseNet-121 on both datasets. Our proposed ThoraX-PriorNet achieves significantly improved scores compared to the vanilla DenseNet-121 using either of abnormality masks. However, we can notice that the disease-prior masks from the VinDr-CXR dataset yield the highest performance in both cases. Especially on the VinDr-CXR test dataset, the improvement for ThoraX-PriorNet is 39.77\% with the VinDr-CXR disease-prior mask, compared to 17.52\% with the NIH chest X-ray14 disease-prior masks. We hypothesize that this is due to two reasons. First, it is due to the quality of the probabilistic maps, as VinDr-CXR has a much higher number of available bounding box annotations. Second, the demographic and class ratio difference between NIH chest X-ray14 and VinDr-CXR may have an effect on the performance. Nevertheless, considering the average improvement in performance compared to vanilla DenseNet-121 with and without aligned images, our proposed model can achieve a significant improvement with either of the disease-prior probabilistic abnormality masks, proving the efficacy of utilizing the APAM block.

\begin{table*}[!t]
\caption{Effect of Random Cropping augmentation on our proposed method with or without test time augmentation.}
\label{table:random crop cls performance}
\centering
\begin{adjustbox}{width=\textwidth}
\begin{threeparttable}
\begin{tabular}{| c | c | c | c | c | c | c | c | c | c | c | c | c | c | c | c | c |}
\hline
TTA & RCrop & Atel & Card & Effu & Infil & Mass & Nodu & Pne1 & Pne2 & Cons & Edem & Emph & Fib & PT & Her & Mean \\
\hline
&  & 81.59 & 89.97 & \textcolor{red}{\textbf{88.27}} & \textcolor{red}{\textbf{72.17}} & 85.02 & 76.74 & 76.16 & \textcolor{red}{\textbf{87.98}} & \textcolor{red}{\textbf{81.63}} & 90.38 & \textcolor{red}{\textbf{92.55}} & 80.74 & 79.52 & 88.98 & 83.69 \\
\hline
& \checkmark & \textcolor{red}{\textbf{82.23}} & \textcolor{red}{\textbf{90.24}} & 88.16 & 71.81 & \textcolor{red}{\textbf{86.05}} & \textcolor{red}{\textbf{77.75}} & \textcolor{red}{\textbf{76.83}} & 87.69 & 81.58 & \textcolor{red}{\textbf{90.83}} & 91.95 & \textcolor{red}{\textbf{81.35}} & \textcolor{red}{\textbf{79.96}} & \textcolor{red}{\textbf{91.89}} & \textcolor{red}{\textbf{84.16}} \\
\hline
TTA & RCrop & Atel & Card & Effu & Infil & Mass & Nodu & Pne1 & Pne2 & Cons & Edem & Emph & Fib & PT & Her & Mean \\
\hline
\checkmark & & 81.74 & 89.68 & 88.00 & \textcolor{red}{\textbf{72.36}} & 84.70 & 77.09 & 76.14 & 87.46 & 81.45 & 90.53 & 92.30 & 79.42 & 79.59 & 90.03 & 83.61 \\
\hline
\checkmark &\checkmark & \textcolor{red}{\textbf{82.54}} & \textcolor{red}{\textbf{90.57}} & \textcolor{red}{\textbf{88.35}} & 72.29 & \textcolor{red}{\textbf{86.39}} & \textcolor{red}{\textbf{78.01}} & \textcolor{red}{\textbf{77.00}} & \textcolor{red}{\textbf{87.96}} & \textcolor{red}{\textbf{81.89}} & \textcolor{red}{\textbf{90.98}} & \textcolor{red}{\textbf{92.38}} & \textcolor{red}{\textbf{81.75}} & \textcolor{red}{\textbf{80.04}} & \textcolor{red}{\textbf{91.90}} & \textcolor{red}{\textbf{84.43}} \\
\hline
\end{tabular}
\begin{tablenotes}
            \item Here, TTA = Test time augmentation, RCrop = Random Cropping, Atel = Atelectasis, Card = Cardiomegaly, Effu = Effusion, Infi = Infiltration, Nodu = Nodule,  Pne1 = Pneumonia, Pne2 = Pneumothorax, Cons = Consolidation, Edem = Edema, Emph = Emphysema, Fib = Fibrosis, PT = Pleural Thickening, Her = Hernia
        \end{tablenotes}
	    \end{threeparttable}
     \end{adjustbox}
\end{table*}

\begin{figure}[!t]
    \centering
    \includegraphics[width=0.9\linewidth]{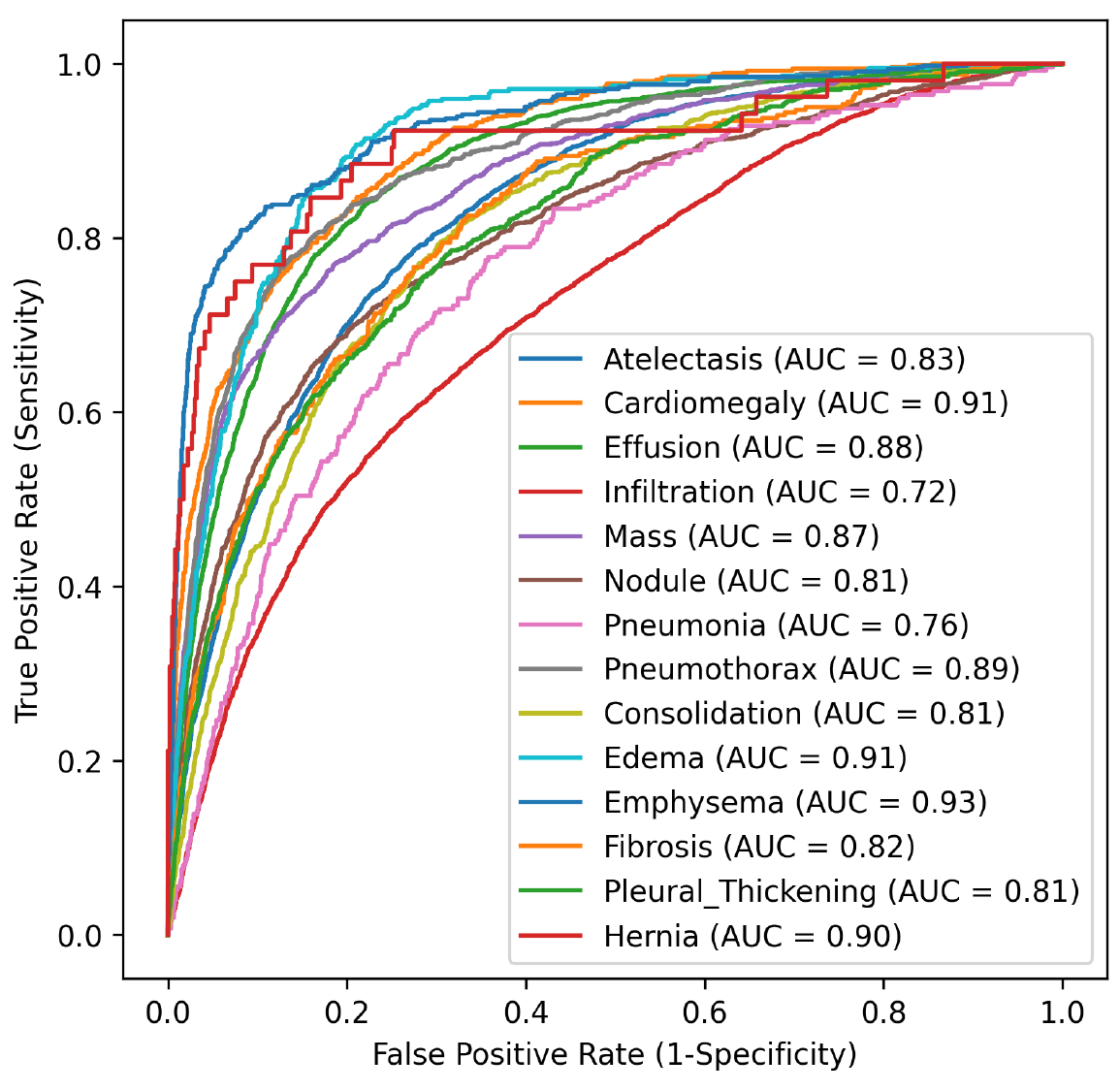}
    \caption{ROC curves of thoracic diseases on the NIH ChestX-Ray14 dataset.}
    \label{fig:roc_curves_nih}
\end{figure}
\copyrightnotice
\subsection{ROC Curves}
The performance of the clinical diagnostic systems is primarily measured by their specificity and sensitivity. The ROC curves are generally used to assess the diagnostic performance of a clinical system by converting the continuous test results into the decision of the presence or absence of pathology and to demonstrate the trade-off between clinical sensitivity and specificity for every possible cut-off for the clinical test. The ROC curves for each pathology on the NIH chest X-ray dataset are shown in Fig. \ref{fig:roc_curves_nih} to visually represent the diagnostic performance of the proposed method.
\copyrightnotice
\subsection{Analysis of Random Cropping Augmentation}
\begin{figure}[!t]
    \centering
    \includegraphics[width=0.9\linewidth]{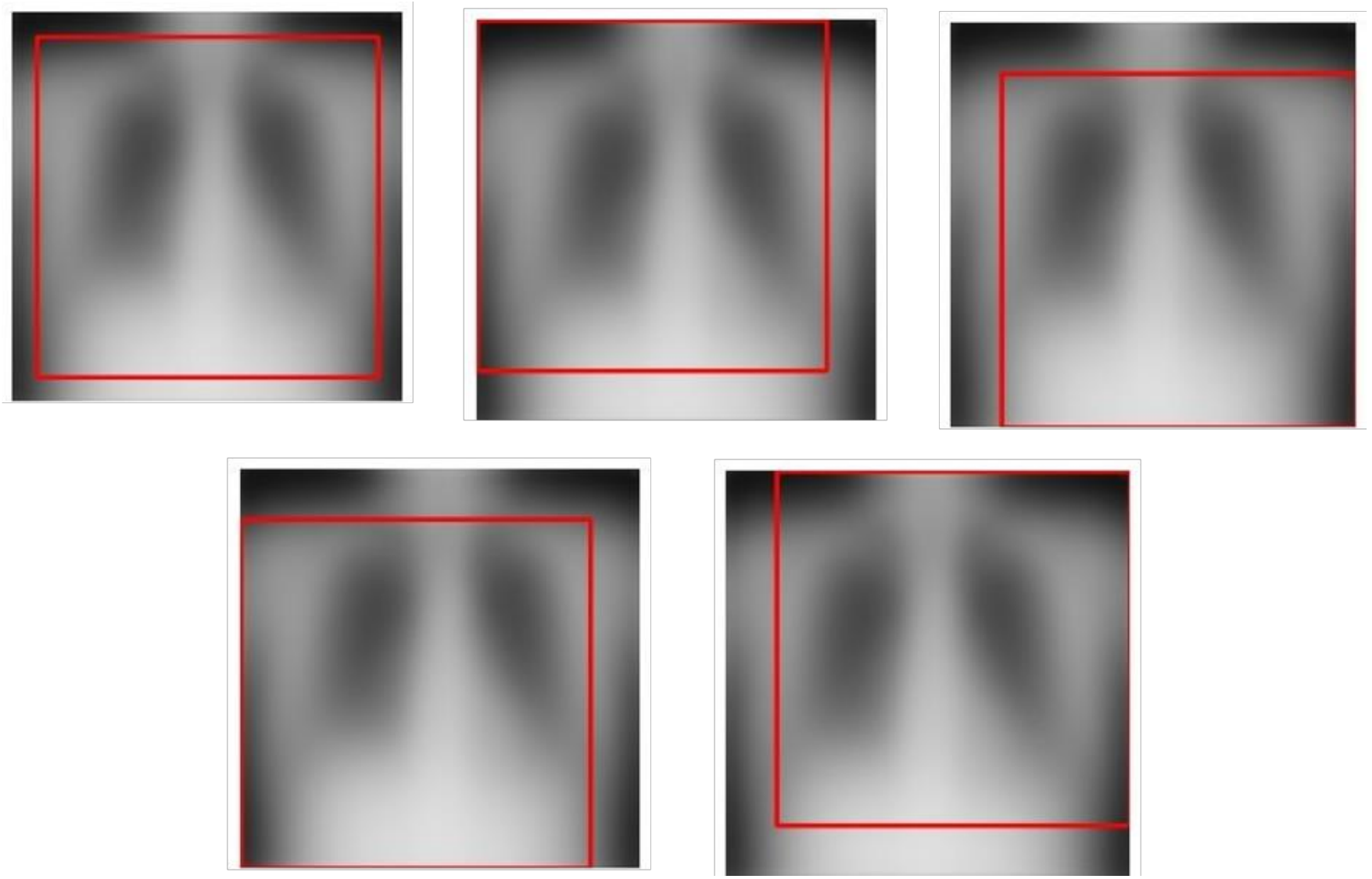}
    \caption{Five different random cropping windows on the anchor image. The red window represents the random cropping window.}
    \label{fig: crop window on anchor image}
\end{figure}

We have utilized the random cropping augmentation following previous studies \cite{Zhu2022PCAN, 9336317}, as the random cropping augmentation has shown improved performance in thoracic disease detection in literature. In addition, we have also performed the alignment of images (where the images are transformed to align their spatial structure with the anchor image \cite{liu2019align}) to ensure that the random cropping technique reliably encompasses all regions of interest within the images. The anchor image is constructed by taking an average of 2000 normal images. In Fig \ref{fig: crop window on anchor image}, we plot five different random cropping windows of size 512\texttimes512 on the anchor image of 586\texttimes586 dimensions (four outmost corners and one centered). We can observe that the random cropping windows can encompass the region of interest.\par
We have also conducted experiments to assess the impact of random cropping augmentation on the performance. The results are reported in Table \ref{table:random crop cls performance}. We can observe that the model performs better when random cropping is utilized. It is intuitive because the random cropping technique significantly augments the training data. Another benefit of utilizing random cropping during training is that we can use test time augmentations (TTA) that consist of different random cropping windows. We have followed the procedure mentioned in \cite{luo2020deep, yan2018weakly} and applied TTA based on random cropping, i.e., utilizing average probabilities of ten cropped sub-images (four corner crops and one central crop and the horizontally flipped version of them) as the final prediction. The results are reported in Table X. We can observe that TTA with random cropping can enhance the performance further.

\subsection{Discussions}
We make several observations by analyzing the extensive experimental evaluation results described in the previous sections. Our studies show that incorporating attention mechanisms like the proposed ThoraX-PriorNet can enhance the performance of thoracic disease classification and localization. The classification accuracy has improved from 84.30\% to 84.67\% for the inclusion of both chest ROI mask and disease-specific mask-based attention in the ThoraX-PriorNet architecture. The improvement in the case of localization is by a more noticeable margin from 0.74 to 0.80 with an IoU threshold of 0.1, 0.56 to 0.63 with an IoU threshold of 0.2, 0.41 to 0.49 with an IoU threshold of 0.3, 0.26 to 0.33 with an IoU threshold of 0.4, 0.14 to 0.22 with an IoU threshold of 0.5, 0.07 to 0.11 with an IoU threshold of 0.6, and 0.03 to 0.04 with an IoU threshold of 0.7. We can also observe that utilizing increased input image spatial resolution or increased feature map dimension shows more notable performance improvement in the localization tasks for diseases with small spatial features (e.g., mass, nodule, pneumothorax).\par
In addition, we have performed the statistical analysis and found the results statistically significant. We have also conducted experiments on the generalizability of the disease-specific prior probabilistic abnormality masks generated from a specific dataset. We observe that though the quality and quantity of the ground truth boxes can affect the generated probabilistic map, our proposed attention mechanism based on the disease-specific probabilistic abnormality masks can achieve superior performance compared to vanilla deep learning architecture.
\copyrightnotice
\section{Conclusion}
\label{sec:Conclusion}
In this work, we present a novel architecture, ThoraX-PriorNet, providing attentions with disease-specific anatomy prior probability maps and chest ROI masks to simultaneously address the CXR image classification and abnormality localization problem. We evaluated our method on two publicly available datasets, NIH ChestX-ray14 and Stanford CheXpert and compared the results with recent state-of-the-art methods.Extensive experiments show that the model, ThoraX-PriorNet performs better by a good margin when considering both classification and localization tasks in a single model and also in the constraint of multiple datasets.

\section*{Acknowledgment}
The authors express their gratitude for the funding provided by the ICT ministry of Bangladesh, which supported this research.

\bibliographystyle{IEEEtran}
\bibliography{main.bib}
\copyrightnotice
\begin{IEEEbiography}[{\includegraphics[width=1in,height=1.25in,clip,keepaspectratio,page=1]{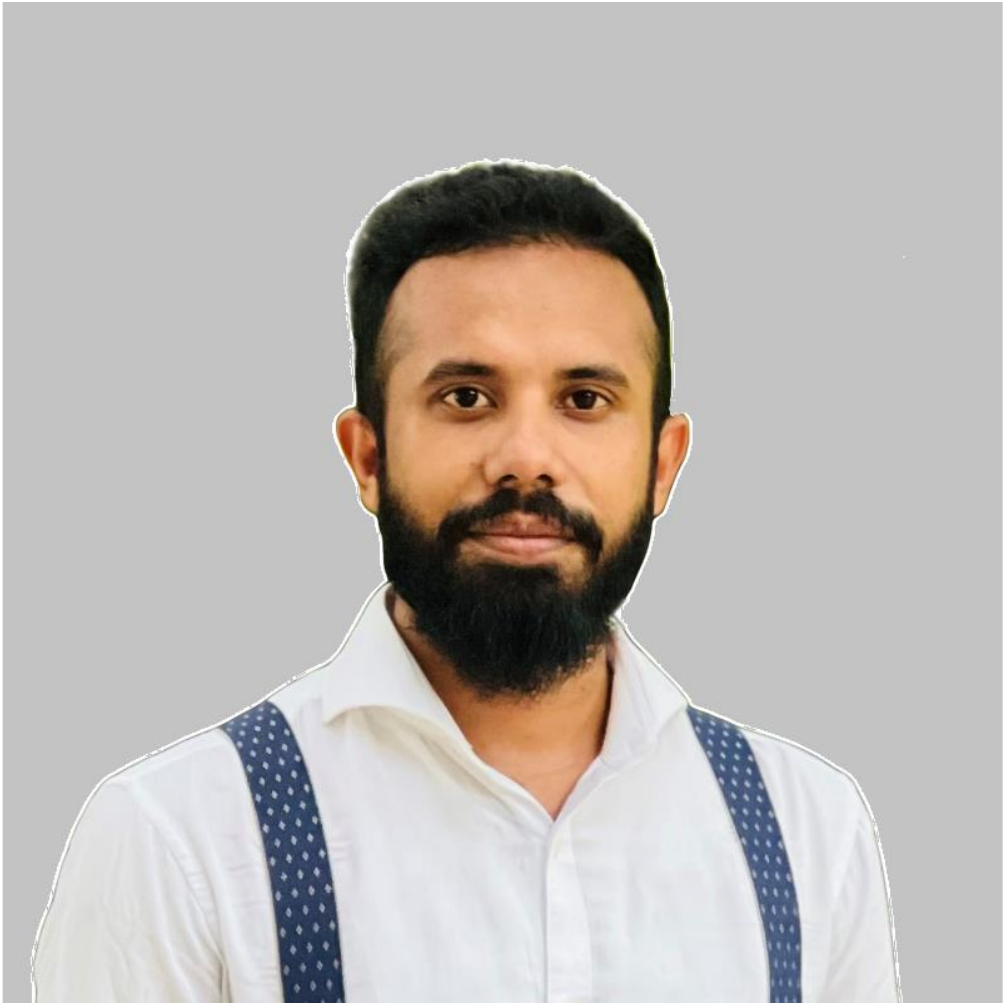}}]{MD. IQBAL HOSSAIN} earned his Bachelor of Science degree in Biomedical Engineering from Bangladesh University of Engineering and Technology (BUET) in 2022. Since mid-2022, he has served as a Research Assistant at the mHealth Lab within the Biomedical Engineering department at BUET, Bangladesh. Subsequently, in 2023, he embarked on his Ph.D. journey in Imaging Science at Washington University in St. Louis. His research focuses on explainable artificial intelligence and medical computer vision.
\end{IEEEbiography}
\begin{IEEEbiography}[{\includegraphics[width=1in,height=1.25in,clip,keepaspectratio,page=2]{figures/author_pics.pdf}}]{Mohammad Zunaed} (Student member, IEEE) completed his B.Sc. and M.Sc. in Electrical and Electronic Engineering from Bangladesh University of Engineering and Technology. He is currently working as a research assistant at the mHealth Lab, Bangladesh University of Engineering and Technology, under the supervision of Dr. Taufiq Hasan. Earlier, he worked as a lecturer in the Electrical and Electronic Engineering department at the Daffodil International University.
\end{IEEEbiography}
\begin{IEEEbiography}[{\includegraphics[width=1in,height=1.25in,clip,keepaspectratio,page=3]{figures/author_pics.pdf}}]{MD. KAWSAR AHMED} received his B.Sc. degree in Biomedical Engineering from Bangladesh University of Engineering and Technology (BUET), Bangladesh, in 2021. He is working as a Lecturer in the Department of Biomedical Engineering, BUET, Bangladesh. His research interests include Machine learning / AI for Biomedical Engineering, Medical imaging, Medical instrumentation \& device design.
\end{IEEEbiography}
\begin{IEEEbiography}[{\includegraphics[width=1in,height=1.25in,clip,keepaspectratio,page=4]{figures/author_pics.pdf}}]{S.M. Jawwad Hossain} received the B.Sc. degree in Biomedical Engineering from 
Bangladesh University of Engineering and Technology (BUET), Bangladesh, in 2022. His research include computer vision and machine learning. 
\end{IEEEbiography}
\begin{IEEEbiography}[{\includegraphics[width=1in,height=1.25in,clip,keepaspectratio,page=5]{figures/author_pics.pdf}}]{Anwarul Hasan} (member, IEEE) Dr Anwarul Hasan is an Associate Professor in the Department of Mechanical and Industrial Engineering, and Biomedical Research Center at Qatar University. Earlier he worked as an Assistant Professor of Biomedical and Mechanical Engineering at the American University of Beirut, Lebanon, and a visiting Assistant Professor and an NSERC Post Doctoral Fellow at the Harvard University and Massachusetts Institute of Technology, USA. Dr Hasan completed his PhD in Mechanical Engineering from University of Alberta, Canada in 2010. His current research interests involve Biomaterials, Tissue Engineering, 3D Bioprinting, Diabetic wound healing, cancer biochips, and Machine Learning and Artificial Intelligence in Health care applications.
\end{IEEEbiography}
\copyrightnotice
\begin{IEEEbiography}[{\includegraphics[width=1in,height=1.25in,clip,keepaspectratio,page=6]{figures/author_pics.pdf}}]{Dr. Taufiq Hasan} (Senior member, IEEE) 
completed his BSc. and MSc. in Electrical and Electronic Engineering (EEE) from Bangladesh University of Engineering and Technology (BUET). He obtained his PhD. in Electrical Engineering from The University of Texas at Dallas, where he  was a member of the Center of Robust Speech Systems (CRSS). He worked as a Research Scientist at Robert Bosch Research and Technology Center, Palo Alto, CA. Dr. Hasan's primary affiliation is with the Department of Biomedical Engineering at BUET as an Associate Professor, where he leads the mHealth research group. He is also affiliated with the Center for Bioengineering Innovation and Design (CBID), Department of Biomedical Engineering at Johns Hopkins University. His research interests are biomedical signal/image analysis, and medical device design. 
\end{IEEEbiography}

\EOD
\balance
\end{document}